# Notes from the Physics Teaching Lab: NMR Experiments at 21 Gauss

Kenneth G. Libbrecht[1]

Department of Physics, California Institute of Technology

**Abstract. We describe a series of laboratory experiments that can be performed with the *Quantum Control* apparatus sold by *TeachSpin*, which uses pulsed NMR techniques to observe the precession of protons in a liquid water sample. With a uniform background magnetic field of 21 Gauss, the protons precess at about 90 kHz, yielding numerous relatively simple observations and measurements that are well suited to undergraduate physics teaching labs. Our goal in this paper is to document some of these experiments in detail, thereby making it easier for instructors to choose material that is best suited for their curricula.**

## Introduction

Nuclear magnetic resonance (NMR) has long been a staple in undergraduate physics teaching labs because the topic checks off so many desirable boxes: 1) the physics incorporates interesting elements from elementary particle properties, statistical mechanics, spin dynamics, and magnetic field interactions, 2) the hardware is robust, relatively inexpensive, and commercially available, and 3) the experiments themselves provide an excellent overall laboratory experience. Recently *TeachSpin Inc.* introduced a new *Quantum Control* NMR apparatus that delivers precision measurement capabilities in an easy-to-use, low-cost instrument. In this paper we document some specific experiments that can be done using this apparatus, with the aim of helping instructors decide what types of investigations fit best into their curricula.

The essential components of the Quantum Control (QC) apparatus are illustrated in Figure 1. A large vertical coil provides a highly uniform background magnetic field of magnitude $B_0 \approx 21$ Gauss, and protons in the Hydrogen nuclei in the water sample precess about this field at about 90 kHz. The $B_0$ field also partially aligns the proton spins, so collectively they produce a net magnetic dipole moment $M_0$ aligned along the vertical $\hat{z}$ axis in thermal equilibrium at room temperature. Starting from this equilibrium, the transverse coil applies a 90 kHz magnetic field perpendicular to $B_0$ that rotates $M_0$ into the $xy$ plane, where it commences precessing about $B_0$. The transverse coil then switches to its second job of detecting the dipole dynamics as its precession induces an emf in the coil. Understanding the physics behind all this and seeing what kinds of measurements can be made with this instrument are the primary goals of this paper.

In this paper we will not attempt to review the detailed physics of NMR, discuss its applications, or survey the field. There are numerous textbooks already available that discuss this branch of physics in great detail ([2001Lev, 2019Blu, 2024Kha]). Nor will we review the many facets of using NMR in undergraduate education, particularly in chemistry where much attention is given to NMR spectroscopy [1982Cal, 2010Blu, 2013Sou, 2015Ben, 2018Hib, 2019Man]. Instead, our focus is mainly on the basic physics of NMR and documenting what one can do in the physics teaching lab with this specific instrument.

---

[1] klibbrecht@gmail.com

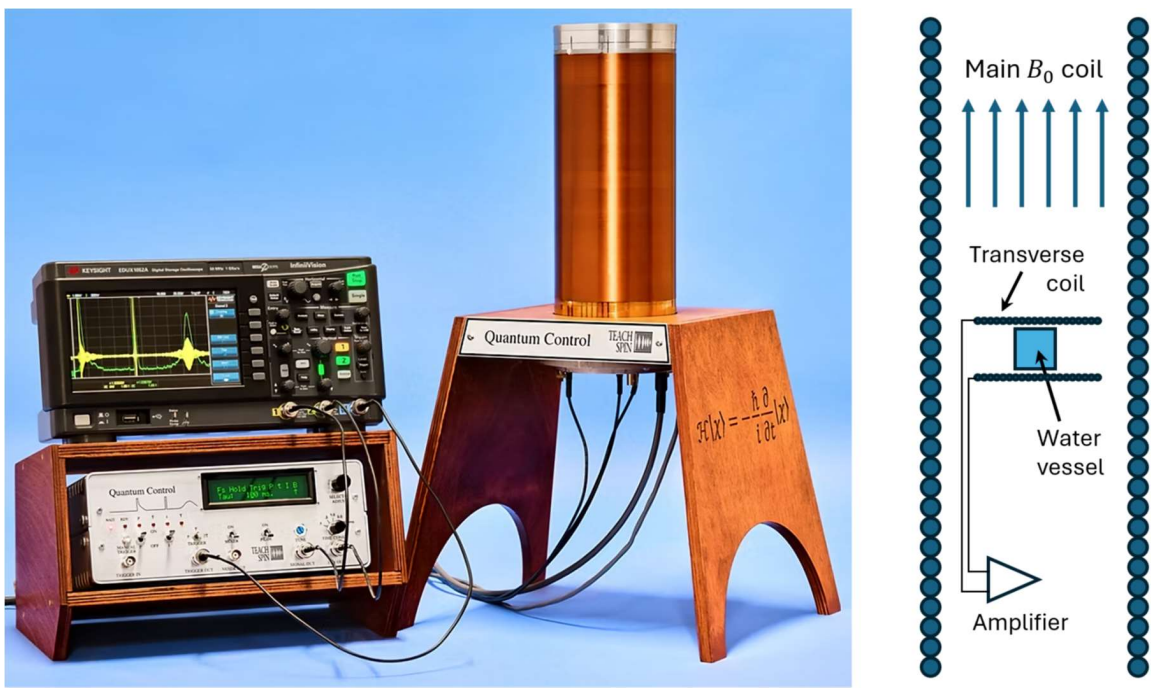

*Figure 1. The TeachSpin Quantum Control (QC) apparatus (left, [2025Tea]) and a schematic diagram of the solenoid coils surrounding the water sample vessel (right). The large vertical $B_0$ coil provides a nearly homogeneous 2.1-mTesla magnetic field in its interior. The transverse coil does double duty: 1) it applies an oscillating magnetic field to rotate the induced collective magnetic dipole of the protons, and 2) it detects its precession of this dipole via induced currents in the coil. The water vessel has no electrical connections, so it only interacts with the QC controller via magnetic fields.*

The problem we are trying to address here is that while it is easy to purchase the QC instrument, it takes a concerted effort to fully understand what it is capable of doing. Of course, the TeachSpin manual is a good resource for the instructor, but it does not provide many examples showing real data and the instrument's full measurement capabilities. Similarly, many laboratory teaching handouts at other universities (available online) tend to describe NMR theory along with a brief look at the specific hardware being used, again leaving new teaching-lab instructors the nontrivial task of deciding what specific laboratory experiences are appropriate for their students.

Unfortunately, instructors often have limited time for exploring the capabilities of new equipment, as this can be a remarkably labor-intensive activity. As a result, we feel that many interesting teaching opportunities may go unrealized, simply because of time constraints. Our aim with this paper, therefore, is to provide the reader with some hopefully valuable information regarding this apparatus and the physics of NMR more generally, so that these notes can be used to inform curriculum choices being made in undergraduate physics teaching labs.

# The induced dipole moment

Our system begins as a collection of essentially free protons (Hydrogen nuclei, only weakly coupled to the surrounding water) sitting in a uniform magnetic field of strength $B_0$. Each proton has an intrinsic magnetic moment $\mu_p = 1.41 \times 10^{-26}$ JT$^{-1}$ and the magnetic field introduces a dipole energy equal to $\Delta E = -\mu_p \cdot B_0$, which depends on the orientation of the proton spin relative to $B_0$ (the latter assumed to be along the z axis). If the protons are all in equilibrium with their surroundings at temperature $T$, then statistical mechanics tells us that the ensemble-average dipole moment per proton will be

$$\langle M_{0,proton} \rangle = \mu_p \tanh\left(\frac{\mu_p B_0}{kT}\right) \approx \mu_p \frac{\mu_p B_0}{kT} \approx 7 \times 10^{-9} \mu_p \tag{1}$$

aligned along $\hat{z}$. Here the numerical value of $7 \times 10^{-9}$ uses $B_0 \approx 2.1$ mT for the QC apparatus and $kT \approx 4.1 \times 10^{-21}$ J at room temperature.

Our 60 mL water sample contains about $(60/18) \times (6.02 \times 10^{23}) = 2.0 \times 10^{24}$ water molecules, so $4.0 \times 10^{24}$ essentially free protons. In thermal equilibrium, therefore, our sample has the induced magnetic moment (counting just the free protons in H nuclei) of

$$M_0 \approx 4 \times 10^{-10} \quad \text{JT}^{-1} \tag{2}$$

Because the number of protons is so large, the net magnetic moment $M$ behaves very much like a simple classical dipole.

While each proton is only slightly polarized by $B_0$, it is often useful to think of the sample as a large "reservoir" of completely unpolarized protons plus a small fraction (about $7 \times 10^{-9}$ in thermal equilibrium) that are perfectly aligned to produce $M_0$. This conceptual picture of separated populations generally yields correct answers in NMR experiments, and it avoids having to think about ensemble averages in statistical mechanics. As we discuss below, this picture can be a useful teaching tool for understanding various NMR signals.

We can also calculate that the value of $M_0$ is relatively free from statistical fluctuations in the QC apparatus. If each proton in a sample of $4 \times 10^{24}$ is polarized by a fraction $7 \times 10^{-9}$, then the total net magnetic moment is the product of these two numbers, so $M_0 \approx 3 \times 10^{16} \mu_p$. For comparison, if we add up the total polarization of $4 \times 10^{24}$ thermally unpolarized protons, then the statistical average is

$$\delta M \approx \left[0 \pm \sqrt{4 \times 10^{24}}\right]\mu_p = [0 \pm (2 \times 10^{12})]\mu_p \tag{3}$$

using standard $\sqrt{N}$ statistics. We conclude that the statistical noise in $M$ from our sample is about $\delta M/M_0 \approx 10^{-4}$, which is quite small. If we wanted to measure the NMR signal from substantially smaller sample (say a few mm³ of human tissue in a Magnetic Resonance Imaging (MRI) machine), then we would need a larger $B_0$ to increase $M_0$ and get a good signal-to-noise ratio.

# Free precession

The next bit of essential NMR physics is the fact that magnetic dipoles precess around magnetic fields, as illustrated in Figure 2. The magnetic field exerts a torque ($\vec{M} \times \vec{B}$) on the dipole, which changes its angular momentum $\vec{L}$ following the equation of motion

$$\frac{d\vec{L}}{dt} = \gamma(\vec{M} \times \vec{B}) \tag{4}$$

where $\gamma = M/L$. For a proton, $L = \hbar/2$ and $\gamma_p$ is a fundamental quantity called the *gyromagnetic ratio* equal to

$$\gamma_p = 2\mu_p/\hbar = 2.675 \times 10^8 \ s^{-1}T^{-1} \tag{5}$$

Thus the equation of motion becomes

$$\frac{d\vec{M}}{dt} = \gamma_p(\vec{M} \times \vec{B}) \tag{6}$$

and solving this gives precession at the angular frequency

$$\omega_p = \gamma_p B \tag{7}$$

and $f_p = \omega_p/2\pi = 42.58$ MHz/Tesla.

In the QC apparatus, the internal oscillator frequency $f_s$ is typically set to match the built-in electronics bandpass filter, and $B_0$ is adjusted to make $f_s \approx f_p$ to high precision. In our particular apparatus, we typically set $f_p = 89300$ Hz, giving $B_0 \approx 2.098$ mT $= 20.98$ Gauss, which occurs with a coil current of about 1.130 A. Thus the main coils give a precession frequency of 42.58 Hz/µT $=$ 79.0 kHz/A.

# Rotating the magnetic dipole

What makes pulsed NMR such a powerful tool is that it is relatively easy to manipulate the proton spins using time-dependent radio-frequency (RF) magnetic fields. More specifically, we manipulate and measure the net magnetic moment $M$, remembering that this is a statistical ensemble of $4 \times 10^{24}$ proton spins. If we do our manipulations and measurements right, then signals from the electrons and other atomic nuclei all average to zero very quickly, allowing us to focus on the proton signal alone. Remarkably, for much of what we will be considering here, the $M$ vector acts much like the simple magnetic dipole illustrated in Figure 2.

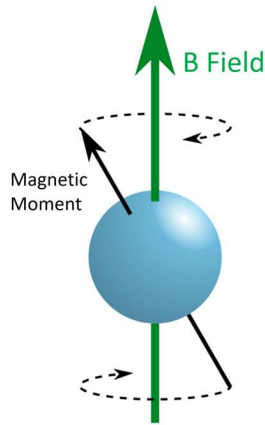

Figure 2. A magnetic moment M will precess around a static magnetic field B. In our NMR apparatus, protons precess about the 2.1 mT (21 Gauss) magnetic field at about 90 kHz. To get the signs right later, note that proton precession about B follows a "left-hand rule": point your left thumb along B and your fingers curl along the precession direction. (Image adapted from https://en.wikipedia.org/wiki/Gyromagnetic_ratio)

Figure 3 illustrates the basic concepts for manipulating $M$ using applied RF magnetic fields in our NMR apparatus, as seen in the laboratory frame of reference. Beginning with the situation in Sketch 1 in the figure, we see $M$ aligned with the static $B_0$ field. As described above, this is what happens when you simply place an ensemble of protons in a static field. This alignment does not happen instantaneously, because the protons are only weakly coupled to the outside world, and some type of "frictional" interaction is needed to reach thermal equilibrium. But if you wait substantially longer than the relaxation time constant of about 3 seconds, then $M$ will "relax" to the value $M_0$ aligned along $B_0$.

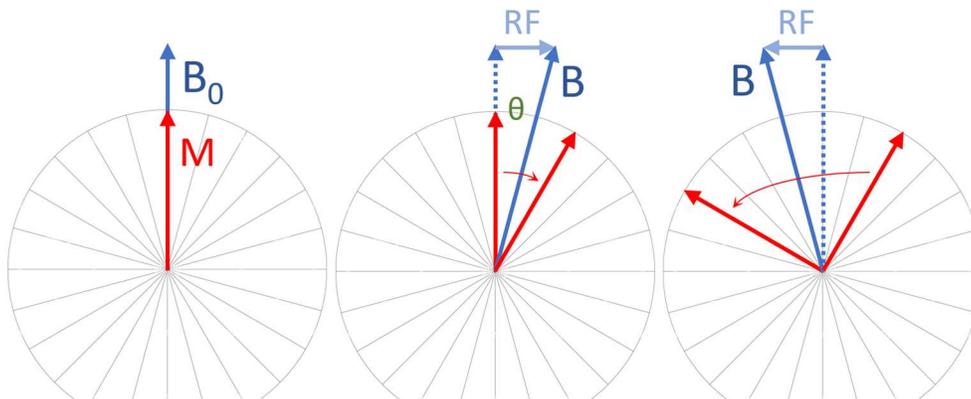

Figure 3. This series of sketches illustrates how RF (radiofrequency) magnetic fields can be used to manipulate an ensemble of proton spins in an NMR sample. Sketch 1 (left) shows the initial equilibrium state. Sketches 2 and 3 show the spins being pushed out of equilibrium by a tailored RF field. In real life, the θ angle from a single RF cycle is much smaller than shown here.

Next apply a small transverse field as shown in Sketch 2, so the vector sum of $B_{RF}$ and $B_0$ gives the net $B$ field shown in the sketch. Keeping $B_{RF}$ static for a bit in this thought experiment, $M$ begins to precess around the combined field. If we wait just long enough for $M$ to precess halfway around $B$, then we have the new situation shown in Sketch 2, where the vertical red arrow has precessed to the second red arrow. Like Sketch 1, Sketch 2 shows only a snapshot in time.

As soon as $M$ has precessed to the position shown in Sketch 2, we then immediately switch the sign of the RF field, giving the new situation shown in Sketch 3. Again we wait half a precession cycle, and $M$ changes direction as shown in the sketch. Now you can imagine repeating the last two steps, causing $M$ to continue its rotation downward. After a single half-cycle of the RF field, $M$ has an angle $2\theta$ from $B_0$, then $4\theta$ after a full cycle, $8\theta$ after two full cycles, etc. In this way, $M$ precesses about $B_0$ while spiraling downward, as illustrated in Figure 4.

To give the qualitative discussion above some mathematical rigor, consider what happens when we start from the equilibrium state with $M = M_0 \hat{z}$ and then apply an oscillating transverse magnetic field that we write as $B_{RF}(t) = B_1 \cos(\omega_p t)\hat{x}$, where $\omega_p$ is set to equal the angular precession frequency of the protons in $B_0$. One way to see what this does is to re-write our applied field as

$$B_{RF}(t) = \frac{B_1}{2}\left[\cos(\omega_p t)\,\hat{x} + \sin(\omega_p t)\,\hat{y}\right] + \frac{B_1}{2}\left[\cos(\omega_p t)\,\hat{x} - \sin(\omega_p t)\,\hat{y}\right] \tag{8}$$

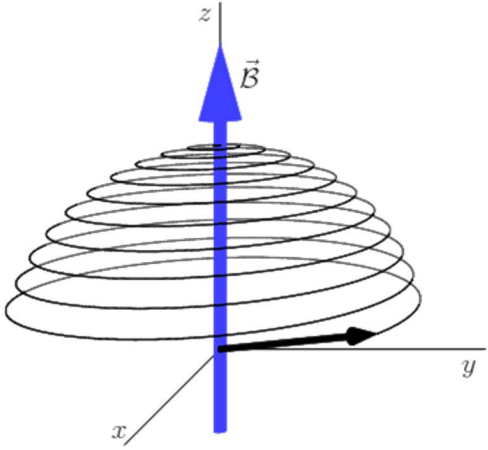

*Figure 4. If we apply a continuous RF field oscillating at the proton precession frequency, the net result is that the net magnetic moment M precesses about the $\hat{z}$-axis while slowly spiraling downward. (Figure from https://web1.eng.famu.fsu.edu/~dommelen/quantum/style_a/nmr.html.)*

The first term in this expression gives a magnetic field vector that rotates with the precessing protons at $\omega_p$, and the second term gives a magnetic field vector that rotates in the opposite direction at $\omega_p$.

If we then switch to a new coordinate system rotating with the precessing protons, we see that our applied field can be written

$$B_{RF}(t) = \frac{B_1}{2}\widehat{x'} + \frac{B_1}{2}\left[\cos(2\omega_p t)\,\widehat{x'} - \sin(2\omega_p t)\,\widehat{y'}\right] \tag{9}$$

Viewed in this rotating frame, the first term gives a *static* field of strength $B_1/2$ with fixed alignment along $\widehat{x'}$, while the second term gives a similar field rotating at $2\omega_p$ in that frame.

The impact of the static field component in this rotating frame is quite simple: the magnetic moment of the protons starts out at $\vec{M} = M_0\hat{z} = M_0\widehat{z'}$, and $\vec{M}$ then executes a slow precession around the $\widehat{x'}$ axis, at a precession frequency equal to $\omega_1 = \gamma_p B_1/2$. If we turn $B_{RF}$ off when the precession angle reaches 90 degrees, this is called a $\pi/2$ pulse. After applying this RF pulse, the magnetic moment in the rotating frame will be static with $\vec{M} = M_0\widehat{y'}$. (Recall that spins precess following the left-hand rule shown in Figure 2.)

Once again, if we watched this process from the non-rotating lab frame, the magnetic moment would execute the rather complex spiral path illustrated in Figure 4. In the rotating frame, however, the dynamics are much simpler to visualize and calculate.

As for the second B-field component rotating at $2\omega_p$ in the rotating frame, it turns out this has a negligible cumulative effect. This rapidly changing field causes the protons to precess a bit, but then quickly precess back again, over and over again. Thus the rapidly rotating field rapidly "shakes" the magnetic moment with no lasting change in $M$. Only the static component in the rotating frame has the cumulative effect of rotating $M$.

Another thing we can see in the rotating frame is that the precession of $\vec{M}$ continues about $\widehat{x'}$ as long as $B_{RF}$ is applied. If we double the $\pi/2$ pulse time, then we get a $\pi$ pulse that rotates $M$ from $+M_0\widehat{z'}$ to $-M_0\widehat{z'}$. If we quadruple the $\pi/2$ pulse time to make a $2\pi$ pulse, then $M$ simply rotates back to its starting point $M_0\widehat{z'}$. We see that theory gives a fairly simple behavior here, although note that it depends on the RF field being applied at precisely the precession frequency. If $\omega_{RF} \neq \omega_p$, then everything quickly gets a lot more complicated.

## The electrical signal from free precession

If we start in equilibrium and apply a $\pi/2$ pulse as just described, then the net magnetic moment $M_0$ will precess around the $\hat{z}$ axis in the lab frame, and we can detect this motion using the transverse coil shown in Figure 1. While calculating the size of the resulting electronic signal with high accuracy is a formidable challenge, it is instructive to make a rough estimate of this signal.

If we approximate our sample as a point dipole with magnetic moment $M$, then the on-axis magnetic field from this dipole is

$$B(r) = \frac{\mu_0}{4\pi} \frac{3M}{r^3} \tag{10}$$

If we place a small loop of wire in this field at a distance $r$, then the total flux through the loop is

$$\Phi \approx \left(\frac{\mu_0}{4\pi} \frac{3M}{r^3}\right) \pi R^2 \tag{11}$$

where $R$ is the radius of the wire loop. Next make a very rough approximation by setting $r \approx R$, giving

$$\Phi \approx \frac{\mu_0}{4} \frac{3M}{R} \approx \frac{\mu_0 M}{R} \tag{12}$$

If we rotate the dipole 180 degrees, the flux through the loop reverses sign, so the resulting EMF is roughly

$$\text{EMF} = -\frac{\Delta\Phi}{\Delta t} \approx \frac{2\mu_0 M}{R\Delta t} \approx \frac{4\mu_0 M f_p}{R} \tag{13}$$

where we have used $\Delta t \approx 1/2f_p$. If we use a pickup coil with $N$ windings, then our induced signal voltage is

$$V_{sig}(t) \approx \frac{4\mu_0 M N f_p}{R} \cos(\omega_p t) \tag{14}$$

oscillating at the precession frequency.

Putting in some numbers (taking $N$ and $R$ from the TeachSpin manual): $\mu_0 = 4\pi \times 10^{-7}$, $N = 400$, $M_0 \approx 4 \times 10^{-10}$ JT$^{-1}$ (calculated above), $f_p \approx 90$ kHz, and $R \approx 2$ cm, we obtain $V_{sig} \approx 4$ µV. Note that $M$ scales as the volume of the sample vessel, so the overall signal scales as $R^2$.

If we make the detector coil part of tuned LC circuit (as described in the TeachSpin manual), then this voltage increases by about a factor of Q, which is the quality factor of the tuned circuit. If we want to observe changes in the free precession signal amplitude that happen over timescales as short as $\Delta t \approx 10/f_p$, then $Q \approx 10$ is optimal, which increases our voltage signal to about $V_{sig} \approx 40$ µV.

To further estimate the signal-to-noise ratio (SNR), we note that a good preamplifier can deliver a noise spectral density of about $V_{nsd} \approx 4$ nV/rtHz (nanovolts per root Hertz). Thus the RMS noise will be

$$V_{rms} \approx V_{nsd}\sqrt{BW} \tag{15}$$

where $BW$ is the measurement bandwidth. TeachSpin uses a tuned circuit to restrict the measurement bandwidth to about $BW \approx 2$ kHz (see Appendix 1) around 90 kHz, thus yielding $V_{rms} \approx 200$ nV and $SNR = V_{sig}/V_{rms} \approx 200$. Although this short calculation describes only a rough estimate of a complex measurement process, a close look at the free-precession signal reveals a SNR of about 100, as shown in Figure 5.

From the standpoint of teaching-lab pedagogy, one can certainly focus on the physical origins of NMR signals without spending any time thinking about how those signals are produced. But understanding electronic signals is certainly an important part of experimental physics and measurement sciences more broadly. Given that the induced emf is described in most first-year physics courses, this NMR instrument provides some excellent teaching moments for connecting theory and experiments in a laboratory setting.

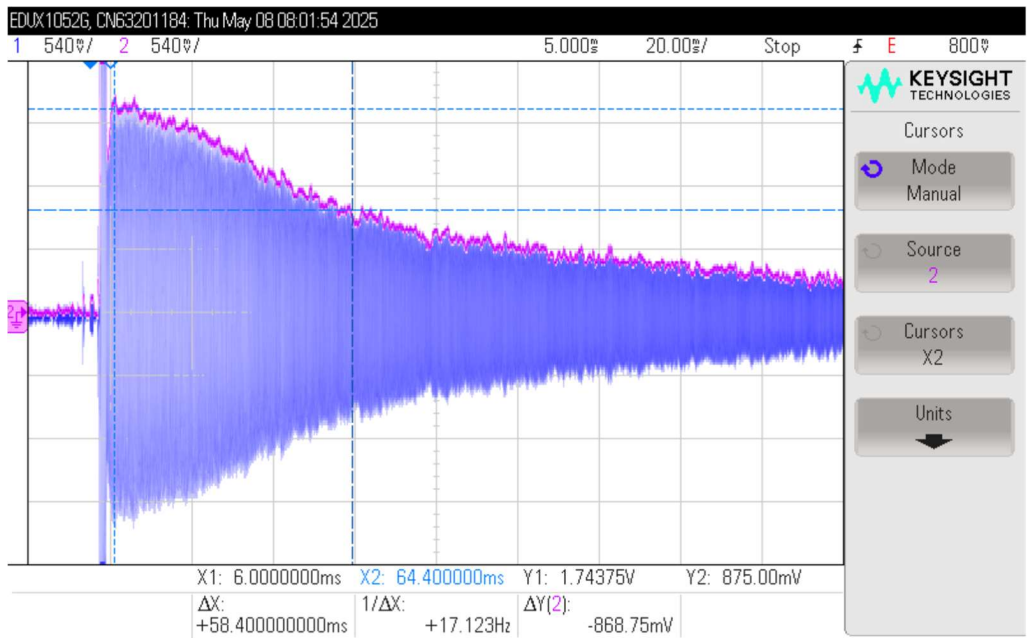

*Figure 5. This oscilloscope screenshot shows a typical free-precession signal (blue) soon after the application of a π/2 pulse, together with the "envelope" signal (red) that shows a running average of the sine-wave amplitude. The free-precession oscillations at 90kHz are too fast to be seen here, while the envelope signal decays with time constant of about 75 msec.*

## Free-precession dephasing

Once a $\pi/2$ pulse has been applied, the protons collectively rotate about $\hat{z}$ at a frequency $f_p$, and the total rotating magnetic moment $M$ induces a current in a pickup coil that is amplified for display on the oscilloscope. If we think of our sample as a collection of fully polarized protons, then at first these particles collectively produce a total magnetic moment $M = M_0$. Over time, however, each proton precesses at a slightly different $f_p$ (owing to B-field nonuniformities), so the free-precession (FP) signal decays away with a time constant of about 75 msec, as illustrated in Figure 5. Put another way, the protons initially precess in a coherent fashion, but the phase coherence is soon lost because of *dephasing*, thus causing a reduction in the FP signal. With a perfectly uniform B-field, the FP signal would last much longer (several seconds), eventually decaying due to coupling between the protons and the surrounding water molecules.

From the observed FP dephasing time, we can estimate the degree of nonuniformity in $B_0$. If half the sample precesses about $B_0$ and the other half precesses about a slightly stronger field $B_0 + b$, then those two samples will dephase at a rate $\omega = d\varphi/dt = \gamma_p b$. Dephasing will reduce the signal substantially when $\varphi \approx 1$ (roughly), giving a free-precession dephasing time of

$$T_2^* \approx \frac{1}{\gamma_p b} \tag{16}$$

where $T_2^*$ usually means the time needed for the FP signal to fall to $(1/e)$ times its initial value. With an observed decay time of $T_2^* \approx 75$ msec, this expression gives $b \approx 50$ nT, which provides a rough estimate of the nonuniformity of the B-field in the water sample vessel. With $B_0 = 2.1$ mT, this means that the field is uniform over the sample to about 25ppm. Much of this small nonuniformity in the field is caused by ferromagnetic objects in the vicinity of the QC apparatus, and these numbers are typical of most laboratory environments. The QC main coil was carefully engineered to produce a B-field uniformity below that found in most laboratory environments.

## Free-precession and the mixer signal

Although the amplitude of the FP signal is the main focus of many of our experimental investigations, one can also glean some information from the FP phase. The raw FP signal can be used to observe the signal phase, of course, but a better approach is to use the "mixer" signal provided by the QC instrument. A mixer is a nonlinear electronic device that essentially multiplies an RF input with a "local oscillator" (LO) input to produce a product signal. In our case the RF input is the FP signal at $f_p$ while the LO input is a sinusoid at the fixed set frequency $f_s$, giving a mixer output

$$V_{mixer} \sim \cos(\omega_s t)\cos(\omega_p t) \tag{17}$$
$$\sim \frac{1}{2}\{\cos[(\omega_p + \omega_s)t] + \cos[(\omega_p - \omega_s)t]\}$$

The QC electronics passes this signal through a low-pass filter to remove the high-frequency term, so the final mixer signal displays a sinusoidal signal oscillating at the difference frequency $f_p - f_s$.

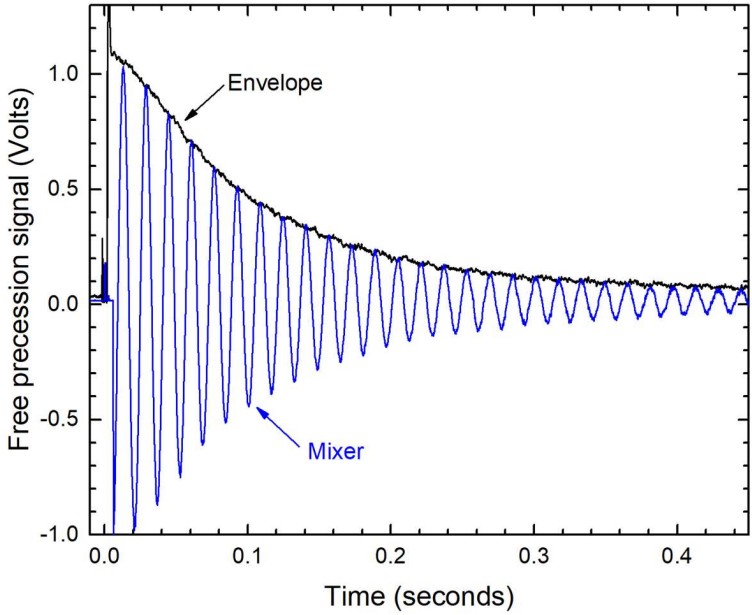

*Figure 6. The mixer signal can be used to provide a nice visualization of the free-precession signal. This plot shows the FP envelope (like that seen in Figure 5) together the mixer signal provided by the QC mixer. The $B_0$ magnetic field was adjusted to give $f_p - f_s \neq 0$, producing a mixer signal that oscillates much more slowly than the raw FP signal.*

One typically uses the mixer signal to adjust $B_0$ to set $f_p - f_s \approx 0$, because this optimizes most of our desired NMR signals. Moreover, NMR theory is much simplified when this difference frequency is set to zero, giving us another good reason to stay at this operating point.

However, the mixer signal can also be used to provide a nice visualization of the FP signal, as shown in Figure 6. Here $B_0$ was adjusted to produce a nonzero difference frequency, allowing us to see the FP signal (reduced to a lower frequency by the mixer) together with the envelope signal on a single data plot. Note that $f_s$ always remains a set constant in the QC apparatus, essentially providing a "clock" with a stable frequency and phase reference.

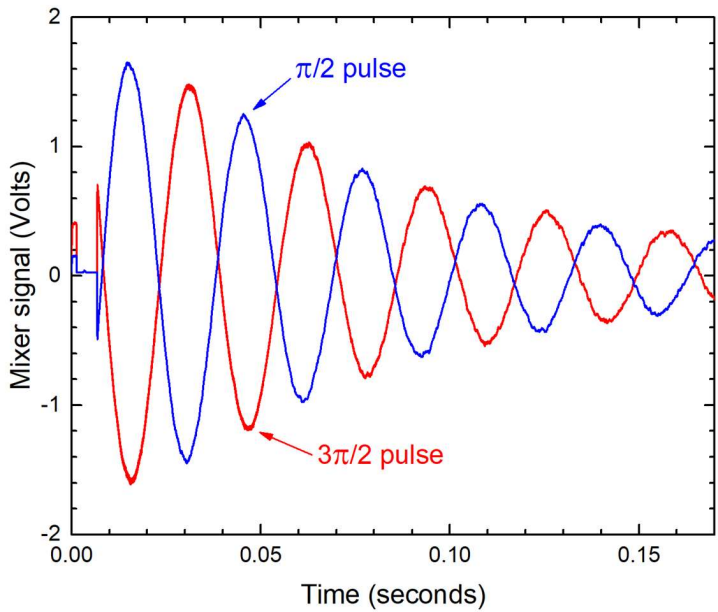

*Figure 7. This superposition of two data sets shows the FP signal (reduced to a lower frequency by the mixer) after a π/2 pulse (blue) and after a 3π/2 pulse. The two pulses were created using the same number of pulse cycles (Ncyc) but different pulse amplitudes. Theory predicts that these two signals should be 180 degrees out of phase, and this is readily observed in the mixer signals (although there is a slight dephasing at longer times).*

Figure 7 shows how the mixer signal can also be used to observe phase changes in the FP signal. From the theory described above, we see that applying a $\pi/2$ pulse rotates the magnetic moment so it points along $+\widehat{y'}$, while applying a $3\pi/2$ pulse yields a magnetic moment aligned along $-\widehat{y'}$. While the envelope signals look identical in these two scenarios, the expected 180-degree phase flip is observed in the mixer signal. Note that the application of any RF pulse train will affect signal phases, so it was important to use the same Ncyc value for the two traces in Figure 7. We therefore changed only the pulse amplitude between traces, leaving all other system parameters the same.

## The Bloch equations

The general behavior of the net magnetic moment $M$ in an NMR system with a uniform applied magnetic field $B_z = B_0$ can be summarized in what are called the *Bloch equations* [1976Kit]:

$$\frac{dM_x}{dt} = \gamma_p (\vec{M} \times \vec{B})_x - \frac{M_x}{T_2}$$

$$\frac{dM_y}{dt} = \gamma_p (\vec{M} \times \vec{B})_y - \frac{M_y}{T_2} \quad (18)$$

$$\frac{dM_z}{dt} = \gamma_p (\vec{M} \times \vec{B})_z - \frac{M_0 - M_z}{T_1}$$

where $\vec{M}$ is the vector magnetic moment at any given time, $\vec{B}$ is the total applied magnetic field, and $T_1$ and $T_2$ are phenomenological "friction" constants that weakly couple the proton spins to the surrounding water. The $\vec{M} \times \vec{B}$ terms here describe the precession of $\vec{M}$ about $\vec{B}$, and much of our discussion above regarding $\pi/2$ and $\pi$ pulses was just a specific application of the Bloch equations.

The time constant $T_1$ describes the thermalization of the magnetic moment along the $z$ axis. Beginning with any initial $M_z$, the magnetic moment typically relaxes to $M_0$ following a simple exponential decay curve. As we will see below, the long-term ensemble average is quite accurately described by the exponential function, reflecting the fact that the number of protons is so large that statistical fluctuations are negligible.

The time constant $T_2$ also arises from a weak coupling with the surrounding water, and this term describes the thermal dephasing of the FP signal resulting solely from these nanoscale interactions. As with $T_1$, the dephasing behavior is also nicely described by an exponential function with a time constant equal to $T_2$. In practice, $T_2$ is usually smaller than $T_1$, although this is not the case for all materials. In pure water at room temperature, $T_1$ is a bit larger than 3 seconds while $T_2$ is a bit smaller than 3 seconds, with some variations depending on temperature, the degree of purity of the water, and perhaps other factors.

It is also customary to define a separate time constant $T_2^*$ that describes dephasing caused by B-field inhomogeneities, as we mentioned above. The dephasing in this case is typically *not* well described by an exponential function, and we describe modeling this type of dephasing behavior below.

The constants $T_1$ and $T_2$ in water can be altered substantially by adding different solutes. Paramagnetic salts, for example, often produce nanoscale magnetic field variations that strongly couple with the otherwise shielded protons, lowering both time constants. In your body, the values of proton $T_1$ and $T_2$ depend on the surrounding tissues, and this is what makes Magnetic Resonance Imaging (MRI) possible. An MRI machine's main job is to measure the proton $T_1$ and $T_2$ as a function of 3D position, and a careful manipulation of these data yields an image of your internal organs. Developing injectable chemical "contrast-enhancing agents" is an area of active research aimed at improving MRI functionality. In a remarkable scientific success story, what began with relatively simple studies of subtle NMR behaviors (like those we can do with the QC apparatus) turned into a substantial industry providing lifesaving medical imaging techniques.

## Short-time signal behavior

The behavior of the QC signals during and immediately after the applied RF pulses can be confusing for newcomers in the teaching lab, especially because the transient signals are extremely large and mostly unrelated to the desired NMR signals we use in most of our investigations. For example, Figure 8 shows the following signal regions:

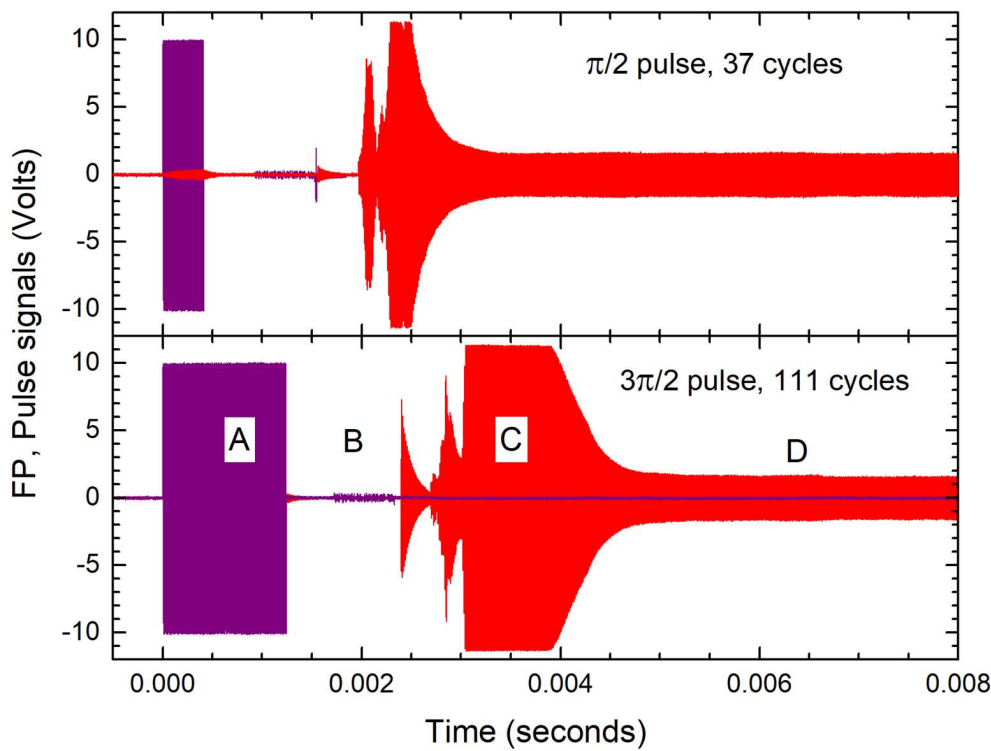

*Figure 8. The two plots above show the application of a 10-volt π/2 pulse (top) and a 10-volt 3π/2 pulse (bottom). The applied RF pulse trains are shown in purple, while the resulting free-precession (FP) signals are shown in red. The lower image shows different signal regions: A) the applied pulses (the RF frequency being so high that individual oscillations cannot be seen in this image), B) a "blanking" region where the FP signal is suppressed, C) a "transient" region where the desired FP signal is strongly contaminated by transient effects, and D) the desired FP signal.*

A. The RF pulse signal is being applied to the sample in the region labeled "A" in Figure 8, with this example showing 37 cycles for a $\pi/2$ pulse and 111 cycles for a $3\pi/2$ pulse, using a pulse amplitude of 10V. Note that a pulse train typically lasts no more than a few msec, and the high-frequency sinusoidal signal can only be seen if one zooms in on this region.
B. In this region the FP signal is suppressed while the transverse coil switches from its two primary uses: supplying the RF pulse and detecting the free-precession signal. The electronics sets the signal voltage to near-zero in this region.
C. Here one sees large voltage transients arising immediately after the detection electronics is activated. Note that the signal often saturates at ±11V in this region, which typically shows a lot of shot-to-shot variability.
D. After the transients die down, a clean FP signal emerges.

Students may be quite distracted by these transient signals when they begin using the QC instrument, because the signal levels are so strong. It is important, therefore, to point out that the desired FP signals mainly occur on longer timescales than seen here.

These transient signals can also have some more subtle detrimental effects on the observed FP signal even after the transients have mostly subsided. In Figure 9, for example, the ±11V transients produced a spurious signal within the oscilloscope's signal processing system. The large transient spike somehow "confuses" the 'scope to produce the signal shown, and we often see this effect in the FP envelope signal when the 'scope gain is turned up especially high. It is easy to work around this problem once you know about it, but it can be an issue when looking for especially low FP signals.

Transient spikes can also distort the FP envelope signal as shown in Figure 10. Once again, this effect is mainly an issue soon after the initial pulse, but that is an important time when making many NMR measurements. Our goal is often to measure FP(0), the magnitude the free-precession signal at $t = 0$, before the FP signal has decayed significantly from $T_2^*$ dephasing. As seen clearly in Figure 10, even fairly minor signal distortions can impede FP(0) measurements. In these cases, using a 0.1-msec time constant is prudent while averaging 'scope traces to improve the SNR.

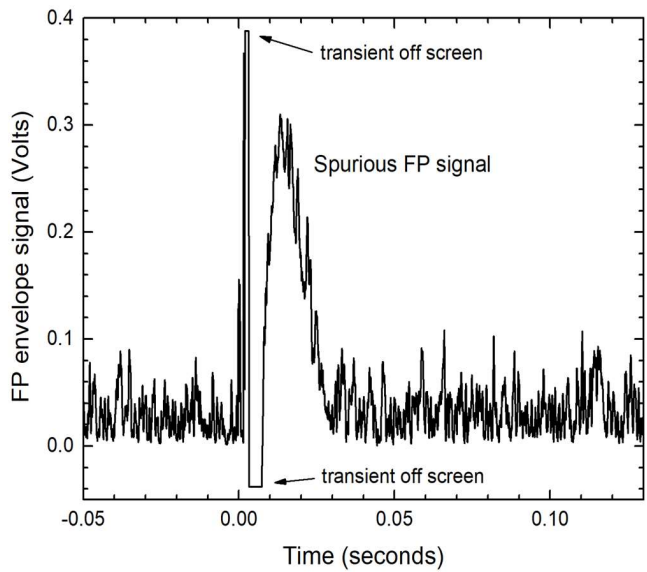

*Figure 9. Many digital oscilloscopes can display "saturation" effects coming from large off-screen spikes. In this example trace, the ±11V transients interacted with the oscilloscope's internal signal processing to produce a sizable spurious signal. The actual FP signal was zero in this example, and we have observed this effect in two different oscilloscopes by different manufacturers. Turning the 'scope gain down usually fixes this problem, but that fix also reduces the signal one wishes to observe.*

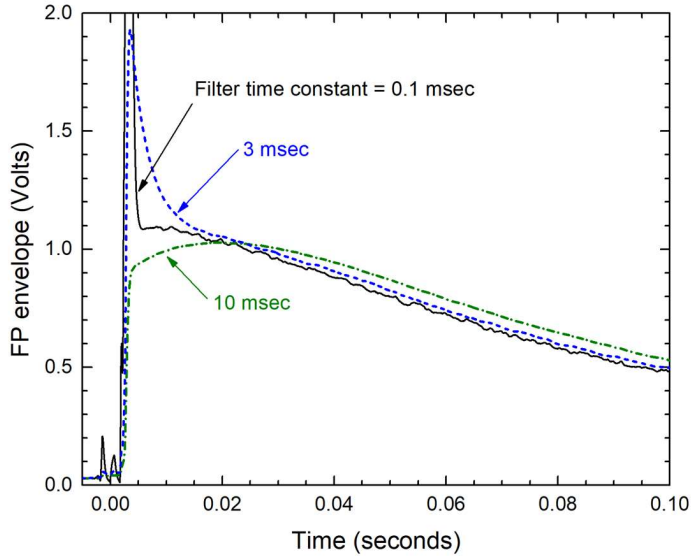

Figure 10. Signal transients can also affect the FP envelope signal when different averaging times are applied. Using the lowest 0.1-msec time constant, the spike only distorts the very beginning of the FP signal. Not surprisingly, the distortion effects are more pronounced when the averaging time is longer, and even the sign of the distortion is not always known.

## Measuring small FP signals

It is also instructive to look at what happens when you try to measure small FP signals using the QC apparatus, because we have found that how one subtracts the background noise becomes an important issue at low signal amplitudes. For example, Figure 11 shows a nice data set looking at FP(0) after a pulse of fixed Ncyc but varying pulse amplitude. Theory predicts a sinusoidal behavior that clearly describes the measurements quite well. However, plotting the same data on a log-log plot illustrates some discrepancies at low amplitudes, as shown in Figure 12.

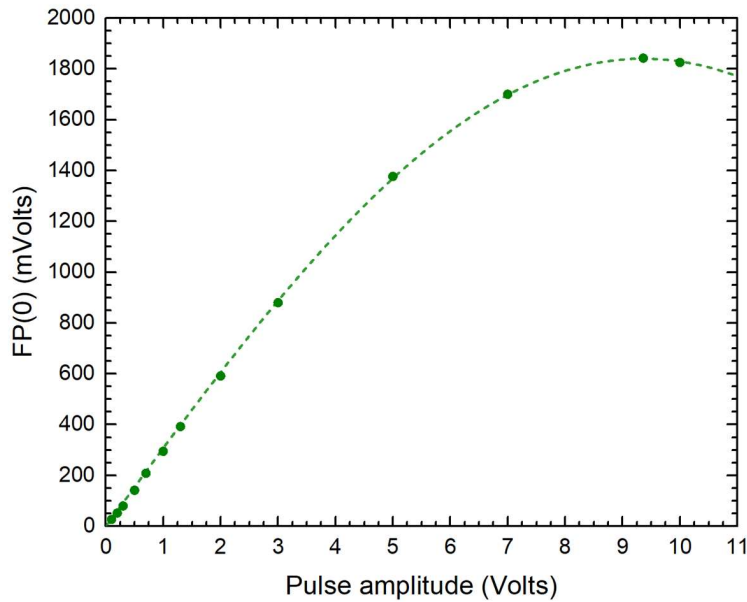

Figure 11. The data points in this plot show measurements of FP(0), the amplitude of the free-precession signal immediately after applying an RF pulse with Ncyc = 40, as a function of the pulse amplitude. With our QC apparatus, this yields a π pulse at Vpulse = 9.36 Volts, and theory predicts the sinusoidal function shown by the dotted line. Clearly the data fit theory quite well.

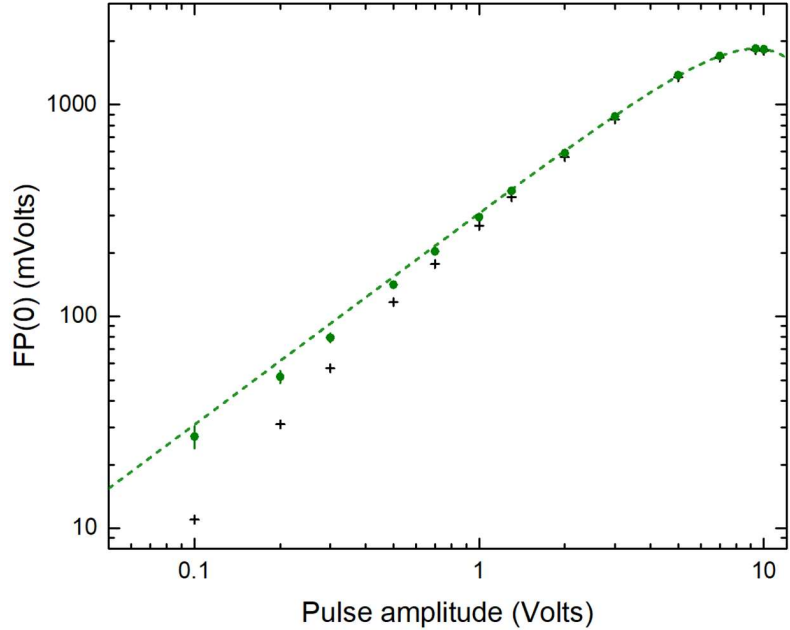

*Figure 12. This plot shows the same data and theory curve as in Figure 11 but in a log-log plot to accentuate the low-amplitude data. The green FP(0) points show the FP(0) after a quadratic subtraction of the background signal described in the text, which is the proper way to remove the background noise. The crosses show the same data but using a direct subtraction of the background signal (also described in the text), which is an improper way to subtract the background. This example shows that the background subtraction can give clearly erroneous results if not done correctly.*

To understand these plots, note that the FP envelope signal has a noise background of around 30 mV typically, and Figure 12 shows two different methods of removing this background. The most obvious method (which students would likely use first) is to simply subtract the background from the signal, giving the corrected measurement $FP_{corr} = FP_{meas} - FP_{noise}$, and this quantity is plotted as the + points in Figure 12.

Choosing this $FP_{corr}$ makes some assumptions about the noise statistics, however, one of which is that the noise should go to zero if many traces are averaged. This is not the case here, because the envelope signal describes the amplitude of the 90 kHz sinusoidal FP signal, and sinusoidal amplitudes are necessarily positive definite. Averaging many traces yields a nonzero value of around 30 mV, and the + points in Figure 12 illustrate that using $FP_{corr}$ gives corrected data that do not match theory at low amplitudes.

In Appendix 1 we describe the noise statistics in detail, showing that (for reasonable assumptions) there is actually no simple analytical form for an optimal subtraction of the noise. Fortunately, a different correction

$$FP_{quad} = \sqrt{FP_{meas}^2 - FP_{noise}^2} \tag{19}$$

is a close approximation of the optimal method, and we plotted $FP_{quad}$ as the green points in Figure 12. These points show much better agreement with theory, plus our noise analysis suggests that $FP_{quad}$ provides a sound correction while $FP_{corr}$ does not.

We see that the low-amplitude $FP_{quad}$ data points in Figure 12 are still systematically below our theoretical expectations, and this could result from a host of other systematic effects in the apparatus or in the NMR system. As described in Appendix 1, our measured noise statistics deviate from pure Gaussian statistics, and there may be additional voltage offsets that are not accounted for. These details notwithstanding, Figure 12 illustrates that the QC apparatus can deliver remarkably precise measurements over a broad range of parameter space, if the noise is handled correctly. While students may not always appreciate these subtler points when making precise measurements, the details can make a difference in one's scientific conclusions.

## Optimizing the pulse amplitude and duration

For many measurements taken with the QC apparatus, one wishes to begin with the equilibrium magnetic moment $M_0$ along the $z$ axis and then rotate it by a known amount, usually via a precise $\pi/2$ or $\pi$ pulse. We have found that the QC apparatus gives quite stable and repeatable results once it has warmed up, so optimizing the pulse properties once is sufficient to obtain accurate data in most circumstances.

There are different ways to determine how the NMR signal responses to different pulses, and Figure 13 shows one example. Here we see that even a sparse data set yields an accurate pulse calibration, which is a testament to the quality of the QC instrument, plus the fact that every proton has precisely the same gyromagnetic ratio. From these and similar data at other voltages, we have found that our specific apparatus delivers an optimal $\pi$ pulse when $N_{cyc} \approx 749.75/V_{pulse}$, where $V_{pulse}$ is the pulse amplitude in volts. With careful data taking, we have found that this formula fits our observations to an accuracy of about 0.3% over a broad range of pulse parameters, as illustrated in Figure 15.

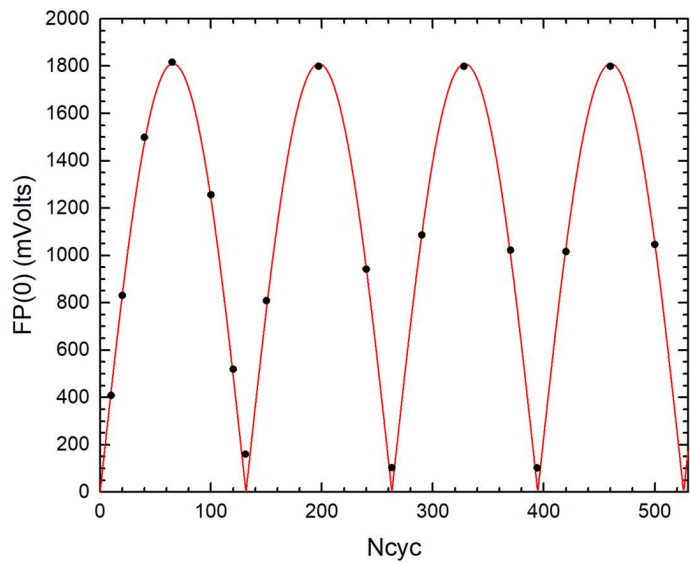

Figure 13. The data points in this plot show measurements of FP(0), the free-precession signal immediately after the application of a rotation pulse with a pulse amplitude of 5.7 volts. The red line shows our model that assumes an optimal π pulse when Ncyc = 749.75/V, where V is the pulse amplitude in volts.

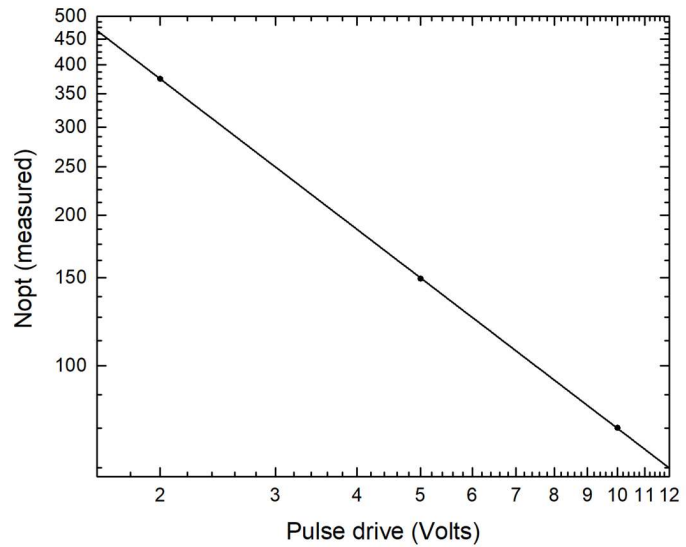

Figure 15. These measurements illustrate a linear model in which an optimal π pulse occurs when Ncyc = 749.75/V, where V is the pulse amplitude in volts. This line fits the measured Nopt (not integer values) to an accuracy of about 0.3%.

We further observed that π pulses generally do not send the FP(0) signal to zero as they would in a perfect world, and that the FP(0) minimum depends substantially on the pulse amplitude. Figure 14 illustrates this by observing FP(0) in the vicinity of a π pulse at different $V_{pulse}$. We do not have a ready model to explain these data, but one possibility is phase jitter during the spin-rotation process. For the data in Figure 14, The optimum $N_{cyc}$ value is about 375, 150, and 75 for pulse amplitudes of 2, 5, and 10 volts, and we would expect that phase coherence would be more of a problem with longer pulse trains. Developing that further into a quantifiable model is a topic for another day.

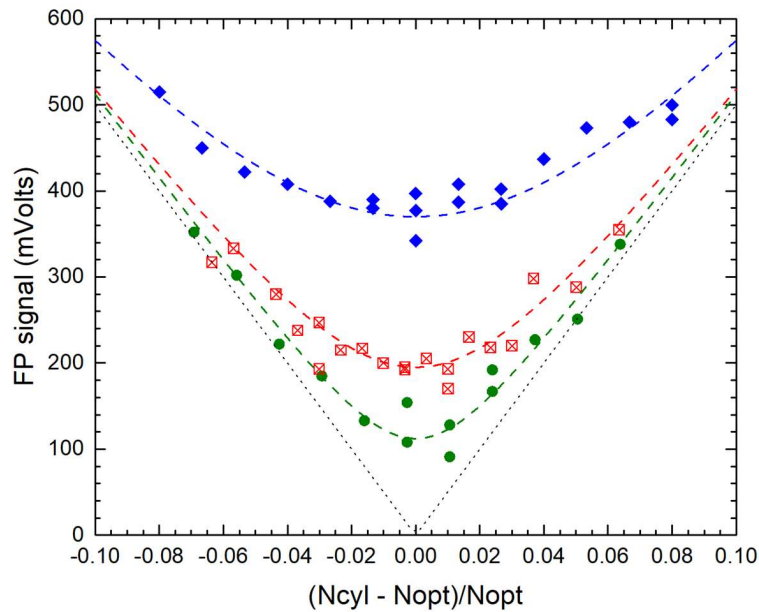

Figure 14. The data points in this plot show FP(0) as a function of Ncyc near a π pulse, using pulse amplitudes of 2V (blue diamonds), 5V (red squares), and 10V (green dots). Nopt is the value of Ncyc that produces an optimal π pulse, and the lower dotted line shows the ideal result in the absence of pulse jitter or other imperfections. The trend illustrates that higher voltages produce π pulses with lower FP(0) minimum values, which is desirable.

We settled on a default set of pulse parameters by setting Ncyc to 80 and observing FP(0) as a function of the pulse amplitude, producing a data set similar to those in Figure 14. With our QC instrument we found Vopt = 9.36 volts, and we now typically set this value whenever we turn the instrument on. Having a convenient integer value of Nopt = 80 for a $\pi$ pulse also has advantages in the teaching lab, just to keep things simple.

## Spin Echo

Observing the spin-echo phenomenon is simple with the QC apparatus, as it was designed with this in mind. Begin by applying the following pulse sequence illustrated in Figure 16:

1) Let the sample equilibrate (H = 5 seconds is enough).
2) Apply a $\pi/2$ pulse at $t = 0$.
3) Wait a time $\tau$, long enough for spins to dephase. ($\tau = 0.5$ seconds is fine)
4) Apply a $\pi$ pulse to rotate the spins 180 degrees about $\hat{x}$.
5) See the spin echo signal after waiting an additional time $\tau$.

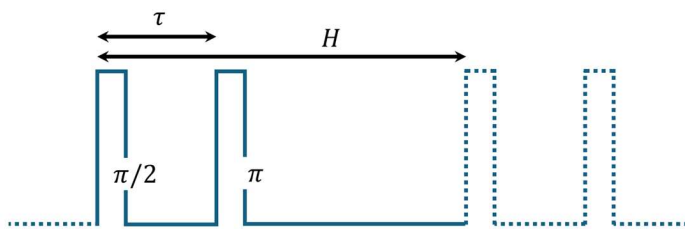

Figure 16. The pulse sequence used to observe the spin-echo signal.

Figure 17 shows the result after averaging traces to clean up the signal. Both the $\pi/2$ and $\pi$ pulses show up as large voltage spikes in this image, mainly from the unwanted short-time transient signals described above. Immediately after $t = 0$ we see the usual free-precession signal, which decays away in a few tenths of a second because of dephasing. Then the $\pi$ pulse reverses the dephasing process, which causes the spins to come back into phase after an additional time $\tau$.

To understand the origin of the spin-echo phenomenon and to characterize its behavior, we can use the Bloch equations to analyze how the magnetic moment evolves throughout the spin-echo sequence. For this analysis and much of what follows, we typically work in the rotating frame, but we drop the primes to keep the notation compact. We begin with a thermalized system having $M = M_0 \hat{z}$ and then apply a $\pi/2$ pulse that rotates the magnetization 90 degrees about $\hat{x}$, yielding a new magnetization $M = M_0 \hat{y}$, as described above. We can analyze this better with matrix notation, using the rotation matrix about the $\hat{x}$ axis

$$R_x(\theta) = \begin{bmatrix} 1 & 0 & 0 \\ 0 & \cos(\theta) & -\sin(\theta) \\ 0 & \sin(\theta) & \cos(\theta) \end{bmatrix} \qquad (20)$$

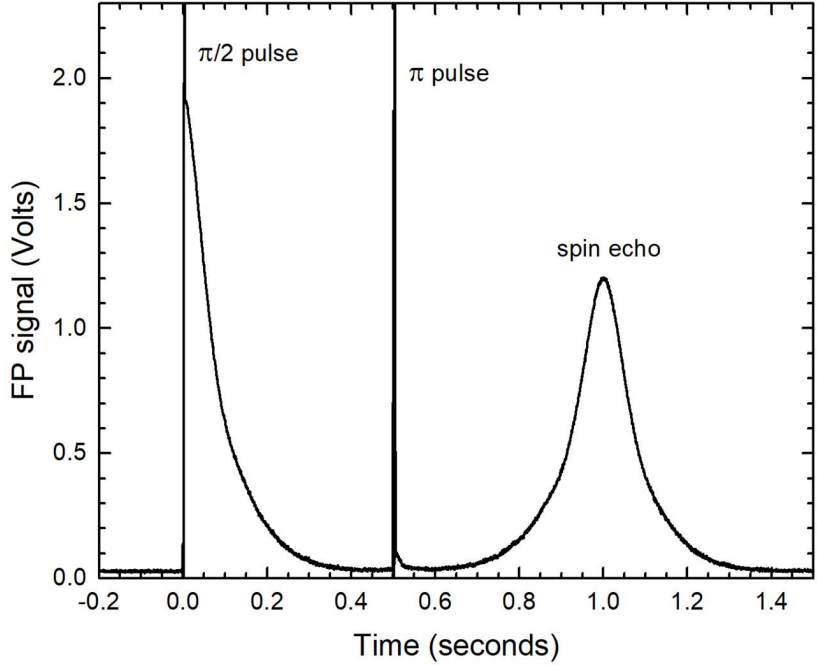

*Figure 17. A basic spin-echo signal with τ = 0.5 seconds, averaging many traces to increase the signal-to-noise ratio.*

A rotation angle of $\theta = -\pi/2$ gives the new magnetic moment vector $M_1 = M_0 \hat{y}$ from

$$M_1 = \begin{bmatrix} 1 & 0 & 0 \\ 0 & \cos(\theta) & -\sin(\theta) \\ 0 & \sin(\theta) & \cos(\theta) \end{bmatrix} \begin{bmatrix} 0 \\ 0 \\ M_0 \end{bmatrix} = \begin{bmatrix} 0 \\ -\sin(\theta) \\ \cos(\theta) \end{bmatrix} M_0 \rightarrow \begin{bmatrix} 0 \\ M_0 \\ 0 \end{bmatrix} \tag{21}$$

We next wait a time $\tau$, during which time the spins precess at slightly different rates about $\hat{z}$ (as the FP signal dephases) until the proton spin vectors are essentially randomly distributed in the $xy$ plane. Using the rotation matrix

$$R_z(\varphi) = \begin{bmatrix} \cos(\varphi) & -\sin(\varphi) & 0 \\ \sin(\varphi) & \cos(\varphi) & 0 \\ 0 & 0 & 1 \end{bmatrix} \tag{22}$$

consider one specific spin that has rotated by an angle $\varphi = \omega_{dp} t$ from its starting position at $\hat{y}$, where $\omega_{dp}$ is the dephasing rotation for that specific spin. Assuming a unit spin, this gives a dephased spin vector

$$S_1 = \begin{bmatrix} \cos(\varphi) & -\sin(\varphi) & 0 \\ \sin(\varphi) & \cos(\varphi) & 0 \\ 0 & 0 & 1 \end{bmatrix} \begin{bmatrix} 0 \\ 1 \\ 0 \end{bmatrix} = \begin{bmatrix} -\sin(\varphi) \\ \cos(\varphi) \\ 0 \end{bmatrix} \tag{23}$$

We next apply our second pulse, leaving the rotation angle $\theta$ arbitrary for now. This rotates this spin vector by an angle $\theta$ about $\hat{x}$, giving the spin vector immediately after the $\theta$-pulse

$$S_2 = \begin{bmatrix} 1 & 0 & 0 \\ 0 & \cos(\theta) & -\sin(\theta) \\ 0 & \sin(\theta) & \cos(\theta) \end{bmatrix} \begin{bmatrix} -\sin(\varphi) \\ \cos(\varphi) \\ 0 \end{bmatrix} = \begin{bmatrix} -\sin(\varphi) \\ \cos(\theta)\cos(\varphi) \\ \sin(\theta)\cos(\varphi) \end{bmatrix} \quad (24)$$

At this point the spin continues to precess around $\hat{z}$ during the second wait time. Assuming the second wait time is also $\tau$, the net rotation angle will again be $\varphi$ from starting point. Thus the spin finally ends up as

$$S_3 = \begin{bmatrix} \cos(\varphi) & -\sin(\varphi) & 0 \\ \sin(\varphi) & \cos(\varphi) & 0 \\ 0 & 0 & 1 \end{bmatrix} \begin{bmatrix} -\sin(\varphi) \\ \cos(\theta)\cos(\varphi) \\ \sin(\theta)\cos(\varphi) \end{bmatrix} = \begin{bmatrix} -\sin(\varphi)\cos(\varphi)[1+\cos(\theta)] \\ -\sin^2(\varphi) + \cos(\theta)\cos^2(\varphi) \\ \sin(\theta)\cos(\varphi) \end{bmatrix}$$

Because all values of $\varphi$ are essentially equally probable after the initial dephasing, we sum up all the spins, essentially averaging $S_3$ over all $\varphi$, to obtain the final spin-echo vector

$$M_2 = \langle S_3 \rangle = \begin{bmatrix} 0 \\ -\frac{1}{2} + \frac{1}{2}\cos(\theta) \\ 0 \end{bmatrix} M_0 = -M_0 \sin^2(\theta/2)\hat{y} \quad (25)$$

Finally, choosing $\theta = -\pi$ for a $\pi$ pulse, we obtain $M_2 = -M_0 \hat{y}$ at the end of the spin-echo sequence.

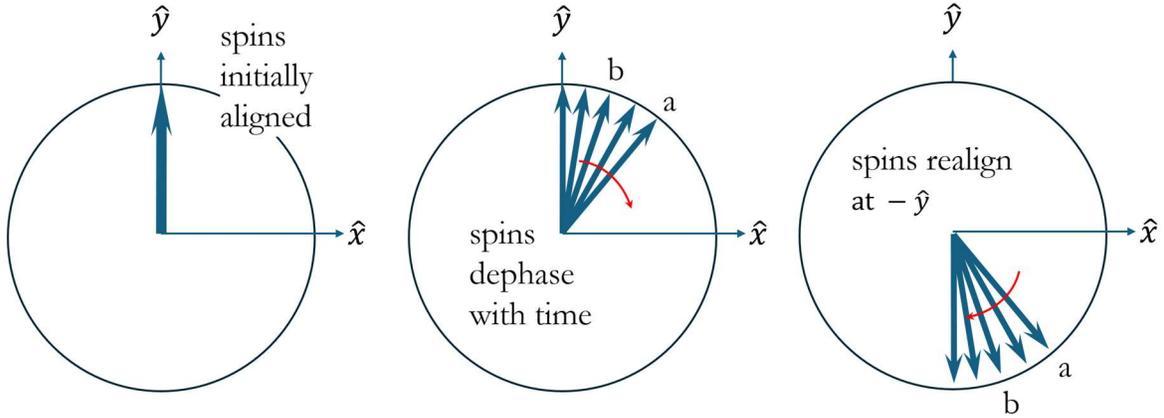

*Figure 18. This series of sketches shows a graphical representation of the spin-echo process. After an initial π/2 pulse, the spins are all aligned along the y axis (left). Over time they dephase from B-field nonuniformities (center). The subsequent π pulse flips the spins about the x axis, at which point the same nonuniformities cause the spins to reunite along the -y axis (right). Note that the spin labeled [a] always precesses faster than spin [b], and that the spins always precess in the same left-hand-rule direction about the z axis.*

Figure 18 provides a graphical illustration of the essential physics of the spin-echo process. Right after the initial $\pi/2$ pulse, the spins are all aligned and collectively they produce a large FP signal. As the spins precess, however, they soon dephase because of B-field nonuniformities, causing the FP signal to rapidly decay to zero. The subsequent $\pi$ pulse then flips the spins so instead of fanning out they reverse and come back into phase after a time $\tau$.

The above analysis ignores $T_2$ damping in the Bloch equations, giving a spin-echo amplitude of $M_0$, equal to the FP signal at $t = 0$. The fact that the actual height of the spin-echo (SE) peak in Figure 17 is lower than the initial FP signal simply reflects the inherent dephasing in the water sample with time constant $T_2$, which we explore in detail below.

Interestingly, the spin-echo phenomenon was not initially predicted from theory but discovered in the lab by accident [1950Hah]. The whole idea of $10^{24}$ proton spins dephasing and then later rephasing was quite counterintuitive at the time. Even now, the whole process seems like some kind of nanoscale magic, making the experiment especially well suited to the undergraduate teaching lab.

We can go further and confirm the result shown in Equation (25) by measuring the height of the SE peak as a function of $\theta$ by varying the length of the second pulse in the sequence. As seen in Figure 19, the data fit the theory curve with remarkable precision.

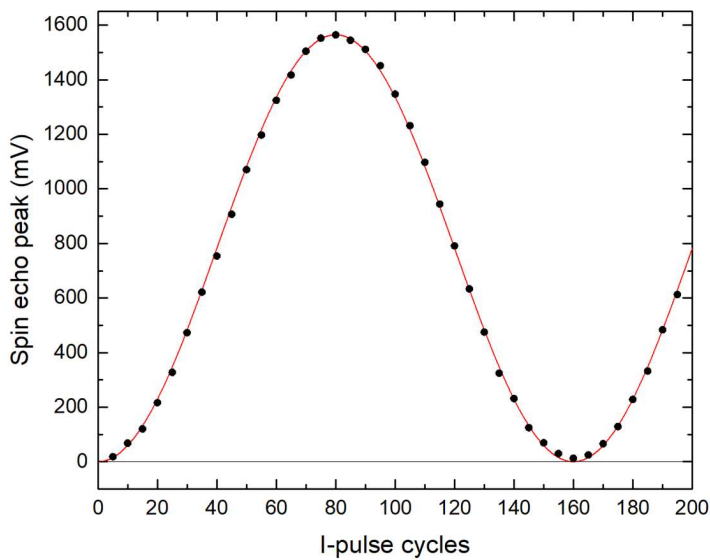

*Figure 19. The data points in this plot show the measured spin-echo (SE) peak as a function of Ncyc for the second pulse, using Ncyc = 40 for the first pulse, with 9.36V for the pulse amplitude and τ = 200 msec. The red theory curve uses only one fit parameter, the peak height.*

## Symmetry of spin-echo signal

If we ignore the $T_2$ relaxation process, theory suggests that the shape of the spin-echo signal should reflect that of the initial FP(t) dephasing. To see this, note that Bloch theory above predicts that the magnetic moment starts out in equilibrium with $M = M_0 \hat{z}$ and rotates to $M = M_0 \hat{y}$ after the initial $\pi/2$ pulse. At a time $\tau$ after the subsequent $\pi$ pulse, the spins realign with $M = -M_0 \hat{y}$ at the spin-echo peak. If we consider the dephasing of the spins immediately after the spin-echo, we see that the dephasing process must be essentially identical to the initial FP dephasing. The starting point is the same (except for a sign, which is irrelevant) and the environment is the same, so the dynamical behavior must be the same as well. Put another way, we can write $SE(t - t_{SEpeak}) \approx FP(t - t_0)$; the spin-echo (SE) signal after the spin-echo peak should be essentially equal to the free-precession dephasing immediately after the initial $\pi/2$ pulse.

Moreover, consider the dynamical behavior of the spins at the spin-echo peak if we run time backwards. The equations of motion do not depend on the direction of time (except for thermal relaxation, which we are ignoring now), and this tells us that the shape of the spin-echo signal should be symmetrical about the spin-echo peak. Thus we expect $SE(-t + t_{SEpeak}) \approx SE(t - t_{SEpeak}) \approx FP(t - t_0)$. Once we know the dephasing behavior of the initial FP signal, we can predict the shape of the entire spin-echo signal as well. Figure 20 provides a nice demonstration of the overall shape and symmetry of the spin-echo signal.

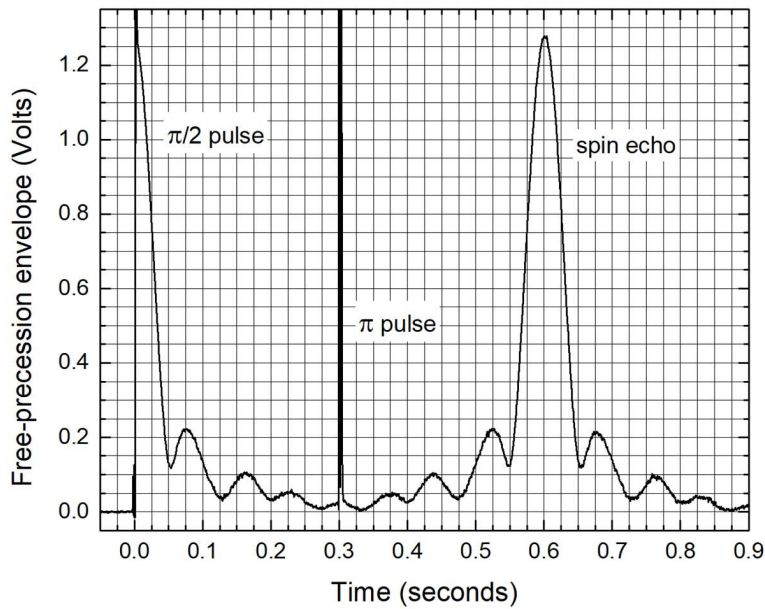

Figure 20. This trace-averaged FP envelope signal demonstrates how the spin-echo signal reflects the initial free-precession decay. We produced this plot by using an external magnetic field to add some structure to the FP dephasing signal (described below), and we multiplied the entire curve by $exp(+(t - t_0)/T_2)$ to correct for irreversible $T_2$ dephasing.

## Phase diffusion

We also note in passing that the spin-echo phenomenon provides a method for observing the phase evolution of the FP signal relative to the QC's local oscillator, as illustrated in Figure 21. Statistical mechanics suggest that the same thermalization mechanism responsible for the $T_2$ spin dephasing should be accompanied by a stochastic phase diffusion, and this process may explain these observations. On the other hand, the phase jitter could just reflect small temporal changes in $B_0$ over these timescales, perhaps from 1/f noise in the coil's current driver. Investigating these effects in detail is another exercise we will leave for a later date. Once again, we see that NMR observations with the QC instrument provide many avenues for further study.

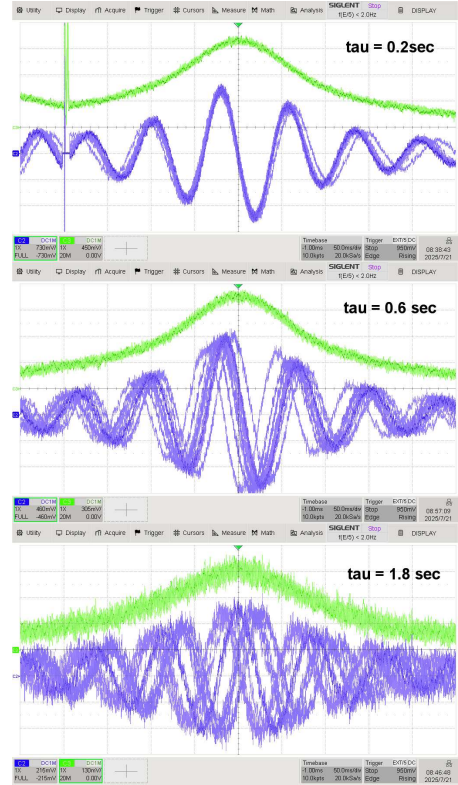

*Figure 21. This series of oscilloscope screen shots illustrates phase jitter than may be related to the $T_2$ relaxation process. The green traces show spin-echo peaks in the FP envelope signal, while the purple traces show the accompanying mixer signals. The magnetic field was adjusted to show slow oscillations in the mixer signals, and 10 traces were combined using the oscilloscope's "persistence" feature. The images show different τ wait times to illustrate the phase noise at later times.*

*These observations reveal that the phase coherence of the free-precession signal decays at roughly the same $T_2$ timescale as the signal amplitude.*

# Measuring $T_1$

The fundamental spin thermalization time $T_1$ can be measured by perturbing the magnetic moment of the spin system and then watching it relax back to $M_0 \hat{z}$. This can be accomplished using different pulse sequences, so let us examine some of the possibilities.

## Method 1 – Consecutive $\pi/2$ pulses

Applying two $\pi/2$ pulses separated in time is the simplest way to measure $T_1$. The first $\pi/2$ pulse rotates $M_0$ into the $xy$ plane, and spin dephasing quickly scrambles the spins to produce $M \approx 0$. Thermalization brings the magnetic moment back to $M_z = M_0$ following

$$M_z(t) = M_0[1 - \exp(-t/T_1)] \qquad (26)$$

If we apply a second $\pi/2$ pulse after a variable wait time, we can measure $M_z(t)$ by measuring FP(0) after the second pulse.

The QC firmware imposes some limitations on pulse timing, so we have to work around these by breaking our data collection process into different pieces. The overall pulse sequence is shown in Figure 22, which is constrained by $H \geq 5$ seconds and $\tau \leq 2.1$ seconds. To produce a full set of data, one can combine data from three slightly different techniques:

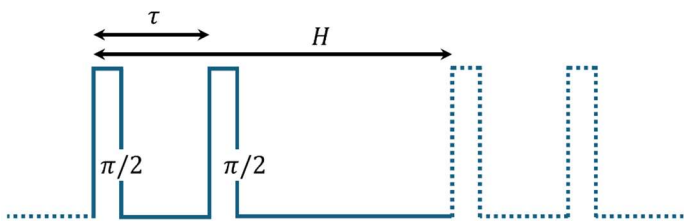

Figure 22. The pulse sequence used to measure $T_1$ using consecutive π/2 pulses (Method 1).

1) Referring to Figure 22, set $H = 5$ seconds, trigger on the second "I" pulse, and measure the free-precession peak as a function of $\tau$. (Ignore the spin echo.) This gives data from 0.3 to 2.1 seconds.
2) Use the same pulse sequence, again with $H = 5$ seconds, but trigger on the initial "P" pulse. Now the I pulse sets M to zero, and the P pulse interrogates M after a delay time of $(H - \tau)$. By varying $\tau$, this gives data from 2.9 to 4.8 seconds.
3) Turn off the I pulse to obtain the pulse sequence (π/2, wait $H$, repeat) and trigger on P. This gives data from 5 to 20 seconds.

Although this three-step process is a bit cumbersome for novice students, plotting all the data together can yield a remarkably nice curve, as illustrated in Figure 23. These data demonstrate that the thermal relaxation process is a simple one, well described by a simple exponential curve with a well-defined $T_1$. This is not surprising, as the measured magnetic moment results from an ensemble average of ~$10^{24}$ water molecules, where essentially every H nucleus experiences identical local conditions.

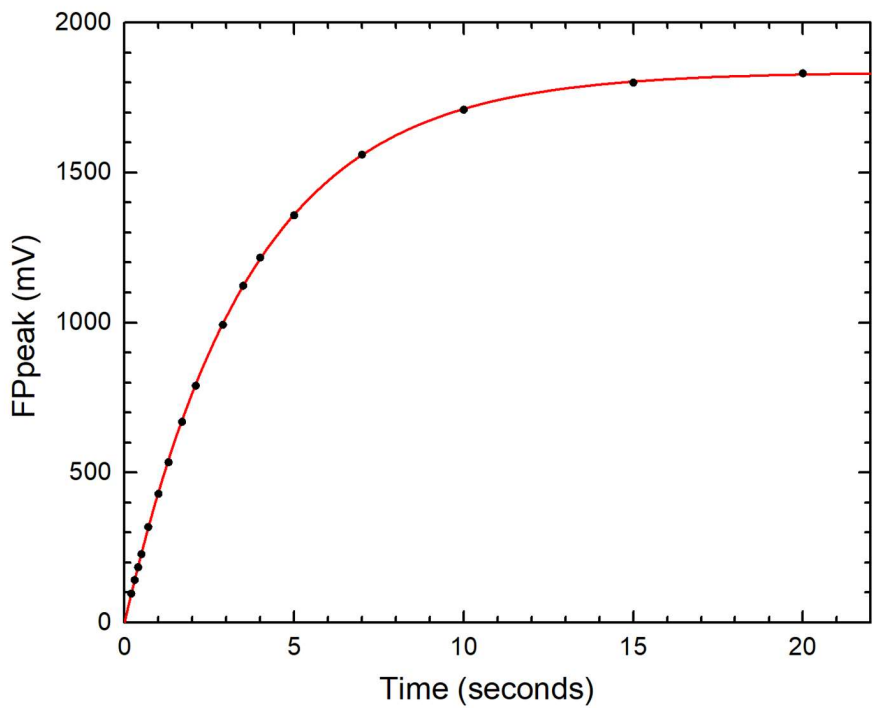

Figure 23. Data points show measurements of the relaxation of $M_z(t)$ using consecutive π/2 pulses (Method 1). The red curve shows $FP(t) = FP_0[1 - \exp(-t/T_1)]$ with $FP_0 = 1835$ mV and $T_1 = 3.7$ seconds.

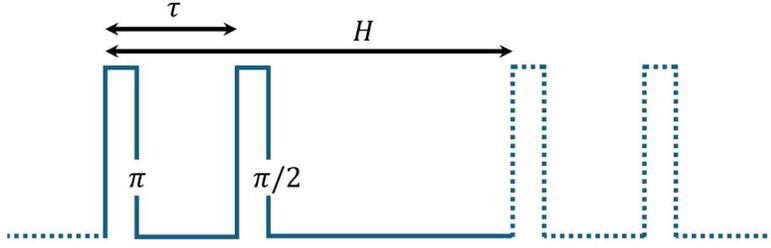

Figure 24. The pulse sequence used to measure $T_1$ using a π pulse followed by a π/2 pulse (Method 2).

## Method 2 – A $\pi$ pulse followed by a $\pi/2$ pulse

Another way to measure $T_1$ is to use the pulse sequence shown in Figure 24), with $H$ set to 20 seconds to return the system to full equilibrium at the end of each cycle. In this case the initial $\pi$ pulse rotates the magnetization vector from $M_z = M_0$ to $M_z = -M_0$, and it subsequently evolves as

$$M_z(\tau) = M_0[1 - 2\exp(-\tau/T_1)] \tag{27}$$

The QC apparatus limits us to $\tau \leq 2.1$ seconds, and Figure 25 show some example data using this pulse sequence. Combining data sets from Method 1 and Method 2 yields the curve shown in Figure 26.

To obtain these data sets, we allowed the QC apparatus to fully warm up, requiring about 24 hours to reach full stability (as we describe below). It is certainly not necessary to wait this long to produce a reasonable measurement of $T_1$, and "student-quality" data can be obtained after much shorter warmup times. But it costs little to turn the apparatus on early, and we have found that the QC instrument can produce extremely high-quality data when fully stable. A full warmup also yields a noticeably longer $T_1$, which increases about 60msec/C with water temperature [2018Tsu, 2024Sie].

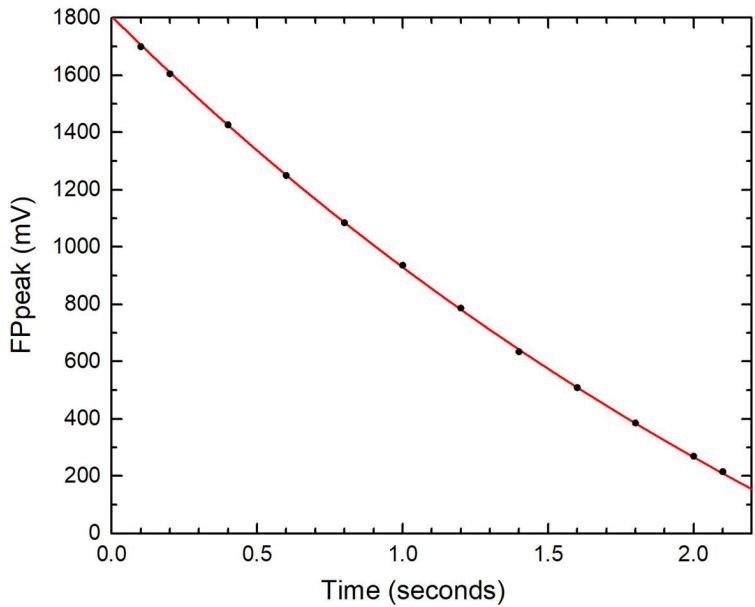

Figure 25. Data points show measurements of M(t) using a π pulse followed by a π/2 pulse (Method 2). The line shows $FP(t) = FP_0[1 - 2\exp(-t/T_1)]$ with $FP_0 = 1805$ mV and $T_1 = 3.6$ seconds.

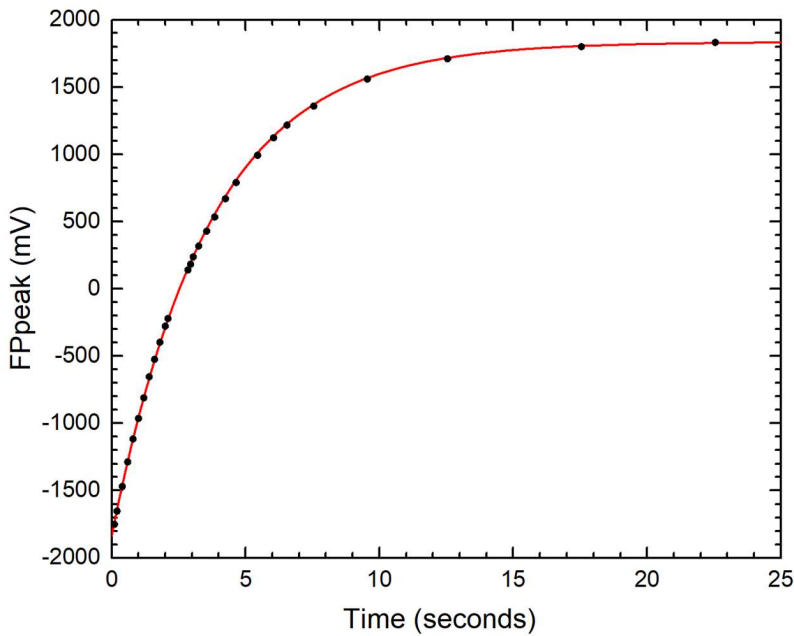

*Figure 26. Reversing the sign of the Method-2 data and combining it with the Method-1 data yields a nice visualization of $M_z(t)$ as it relaxes from $-M_0$ to $+M_0$ after an initial π pulse.*

# Adding external magnetic fields

It can be quite interesting and educational to modify the free-precession signal using externally applied magnetic fields, both using permanent magnets and external magnetic coils placed near the QC apparatus. With permanent magnets, one can easily observe the effects of B-field nonuniformities on the FP dephasing simply by waving a magnet around in real time, although a minimum H time of five seconds is not optimal for student engagement. Cell phones may have noticeable effects as well, if they are brought close enough. With a stationary magnetic coil, the FP signal changes are more reproducible, and can even be modeled with some success, as we discuss below.

We have been using a 15-cm-diameter coil for this purpose, which simply rests on the table with the QC coil apparatus, directly underneath the main QC coil. The coil is thus about 40 cm below the sample vessel, with the coil axis being vertical. Depending on the ambient magnetic fields in the laboratory, we have found that the FP signal may vary somewhat when the coil is moved around on the table or when the current polarity is reversed. To lowest order, however, we can model the coil effects to a reasonable degree by assuming a simple wire coil, a constant internal $B_0$ produced by the main QC coil, and negligible laboratory background fields.

We calibrate the B-field produced with this coil by observing the mixer signal. For our particular coil, we found that a current of 0.1A changed the FP frequency by 88 Hz, giving a mean field at the water sample of about $B_1 = 20.7$ µT/A. With a coil resistance of 21 Ohm, it is sufficient to drive the coil with a function generator, making it easy to change the DC polarity, amplitude, and time-dependence of the applied field.

Along the coil axis of a simple coil, theory tells us that the magnetic field is given by

$$B_1(z) = \frac{\mu_0 N I R^2}{2(R^2 + z^2)^{3/2}} \tag{28}$$

where $R$ is the coil radius, $I$ is the coil current, $z$ is the axial distance, and $N$ is the number of wire turns in the coil. Thus we can write the field gradient as

$$\frac{dB_1}{dz}(z) = \frac{3z}{R^2 + z^2} B_1(z) \tag{29}$$

and using $z = 40$ cm and $R = 7.5$ cm yields $dB_1/dz \approx 150$ µT/A-m. By calibrating using the mixer signal, one does not need to know the number of turns, allowing us (and you) to use any coil you happen to have lying around the lab.

We used this coil to produce the FP signal shown in Figure 20, as a simple field gradient can produce an oscillatory behavior in the FP decay. When a simple field gradient is desired (which is usually the case) the $B_0$ field is adjusted to produce a near-constant mixer signal, thus canceling the constant-field contribution from the coil.

In addition, we often find external fields useful for hastening the FP dephasing. As illustrated in Figure 27, this method makes it possible to better observe FP signals at low $\tau$ values. External magnetic fields were used in this way to give the low-$\tau$ data points in Figure 23, for example.

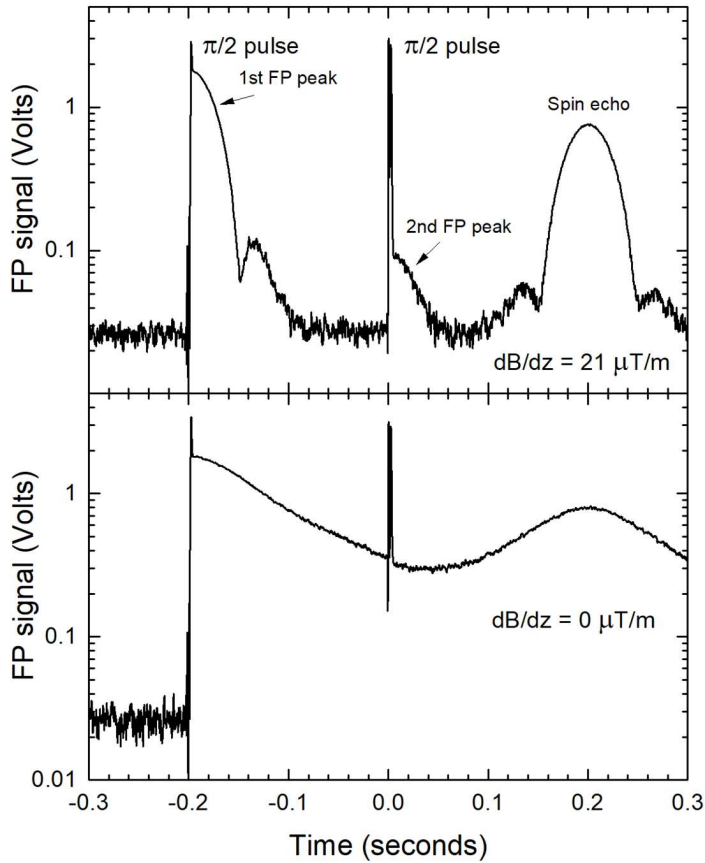

*Figure 27. These data show a trick one can use to measure FP signals when the time delay is shorter than the normal FP dephasing time. The top log plot shows a trace-averaged FP(t) signal during two consecutive π/2 pulses separated by 0.2 seconds. Here an applied field gradient has reduced the dephasing time, so the small second FP peak is clearly visible and measurable. The short-time data in Figure 23 were obtained using this method.*

*The bottom plot shows essentially the same FP signal but without the applied field gradient. Here the initial FP signal has not had time to dephase before the second π/2 pulse was applied, producing a signal confusion that buries the second FP peak.*

# Modeling FP dephasing

In this section we attempt to model the free-precession dephasing signal to see what we can learn about the underlying physics along the way. Our overarching goal is to observe $FP(t)$ under a variety of conditions, including in the presence of applied external magnetic fields, and build a mathematical model that reproduces the observations as best we can. While this topic is somewhat advanced, it examines one of the most obvious signals seen with the QC instrument. The theory also provides a nice application of Fourier methods, so may be of interest to some lab instructors.

We begin by splitting our water sample into $N$ independent voxels (volumetric pixels), each possessing a magnetic moment $M_0/N$ in equilibrium. Starting from thermal equilibrium, we next apply a $\pi/2$ pulse that leaves $M$ in each voxel precisely aligned along the $\hat{y}$ axis at $t=0$. Because the magnetic field is slightly inhomogeneous, each $M_i$ precesses at a slightly different rate. To calculate $FP(t)$, therefore, we must sum up the contributions of each voxel to determine the measured signal.

Conceptually, this "mesoscopic-average" approach allows us to think of these voxels as if they were essentially identical spins, each possessing a finite magnetic moment that is free from statistical fluctuations. If instead we dealt with individual protons, then we would have to address the fact that each small magnetic moment fluctuates wildly from thermal motion. Our voxels are small enough that each sits in an effectively uniform magnetic field, but large enough to have a well-defined magnetic moment. To simplify the nomenclature, we will refer to each voxel magnetic moment as a "spin."

We next write the magnetic field within the water sample as

$$B(x) = B_0 + b(x) \tag{30}$$

where $B_0$ is a constant independent of 3D spatial position $x$ and $b(x)$ is the perturbation field. We will treat $B(x)$ as a scalar because $b \ll B_0$ and what matters for the FP dephasing is mainly the overall rate of precession at each point in the sample vessel. For this reason, components of $b(x)$ along the $z$ axis contribute linearly to changing the precession frequency, while transverse components are reduced by a factor of $b/B_0 \ll 1$, yielding only small quadratic effects in our dephasing model. This can easily be demonstrated in the lab: applying a nonuniform external magnetic field along the $z$ direction has a large effect on the dephasing signal (and the mixer signal), while transverse applied fields do not.

With this definition of the field nonuniformities, the illustration in Figure 28 shows how the spins "fan out" in the rotating frame, which we define by precession in a $B_0$ field. This fanning-out is responsible for the FP dephasing, so this process is what we need to describe in our quantitative model.

We next define a distribution function $D(b) = dN/db$ that derives from the perturbation field throughout the sample. For each slice of the perturbation field with values between $b$ and $b+db$, $dN = D(b)db$ gives the number of spins experiencing fields within this range. Adding up all the spin vectors in Figure 28 then gives our desired signal

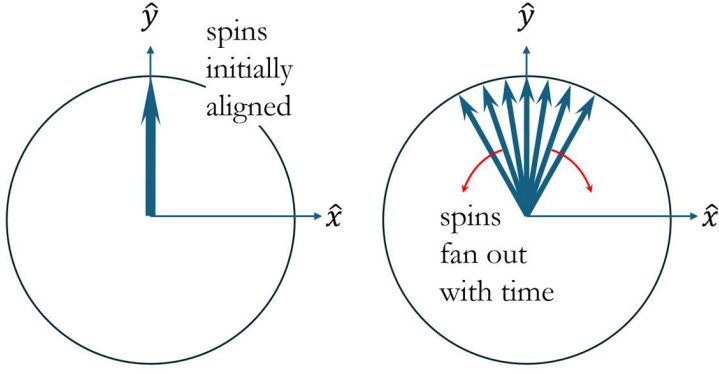

Figure 28. In our model system, the spins start out aligned in the rotating frame at t=0 (left). The spins subsequently lose this initial alignment as each precesses at a slightly different rate (right). Here we have assumed a symmetrical distribution function, as defined in the text.

$$FP(t) = \int_{b_{min}}^{b_{max}} D(b) \exp(i\gamma_p bt)\, db \tag{31}$$

where $\omega_p = \gamma_p b$ is the voxel angular precession frequency. In this complex expression, the real and imaginary parts of $FP(t)$ give its components along the $\hat{y}$ and $\hat{x}$ directions, respectively.

When observing $FP(t)$, we typically adjust $B_0$ to produce a nearly constant mixer signal, which means that the weighted average $\langle b \rangle$ is essentially zero. In what follows, we often assume a "symmetrical" model with $D(-b) = D(b)$, as this assumption automatically gives $\langle b \rangle = 0$. In this symmetrical case Equation (31) reduces to the real expression

$$FP(t) = \int_0^{b_{max}} D(b) \cos(\gamma_p bt)\, db \tag{32}$$

and the component of $FP(t)$ along the $\hat{x}$ axis is always zero, as illustrated in Figure 28. (Note that we are ignoring changes in the normalization of $D(b)$ at this point, as supplying a proper normalization is easily done at the end of the analysis.) The actual distribution function need not be symmetrical, but making this assumption provides a simpler first analysis of the problem.

Defining $a = b/b_0$ and $\xi = \gamma_p b_0 t$ gives the dimensionless form

$$FP(t) = \int_0^{a_{max}} D(a) \cos(a\xi)\, da \tag{33}$$

where $b_0$ is some constant field value associated with the specific model being used. If we transform back from the rotating frame to the lab frame, then the full FP signal is $FP_{full}(t) = FP(t)\cos(\omega_{p,0}t)$, where $FP(t)$ is a real function above and $\omega_{p,0} = \gamma_p B_0$. In a symmetrical model, $FP_{full}(t)$ can be either in phase with $\cos(\omega_{p,0}t)$ or 180 degrees out of phase; no other phase options are possible. If all we observe is the envelope signal (which contains no phase information), then

$$FP_{envelope}(t) = |FP(t)| \tag{34}$$

In principle, one could use Fourier-transform techniques to invert this model and produce a model of $D(a)$ directly from the measured $FP(t)$, but this approach tends to obscure the underlying physics. Instead, it is instructive to begin by examining some special cases.

## A Gaussian model

It is immediately tempting to consider a model in which the B-field "noise" field $b(x)$ could be approximated by a Gaussian distribution, and the math certainly works out nicely with this functional form. Choosing $a = b/b_{RMS}$ and defining $D(a)$ as

$$D(a) \sim \exp\left[-\frac{a^2}{2}\right] \qquad (35)$$

gives

$$FP(t) \sim \exp\left[-\frac{(\gamma_p b_{RMS} t)^2}{2}\right] \qquad (36)$$

in the limit that $a_{max} \gg 1$. Defining $T_2^*$ by $FP(T_2^*) = 1/e$ then gives $\gamma_p b_{RMS} T_2^* = \sqrt{2}$. This simple analytic expression essentially results from the fact that the Fourier transform of a Gaussian is another Gaussian.

While this is a reasonable starting point for conceptual reasons, there is no compelling reason to expect that a Gaussian model would fit the FP data, because the B-field distribution in the sample vessel does not result from any type of uncorrelated Gaussian noise. Indeed, we have found that the Gaussian model generally provides a poor representation of the data.

## An exponential model

Another well-known Fourier-transform pair involves the exponential and Lorentzian functions, so we can try a model with $a = b/b_0$ and defining $D(a)$ as

$$D(a) \sim e^{-|a|} \qquad (37)$$

which gives

$$FP(t) \sim \frac{1}{1 + (\gamma_p b_0 t)^2} \qquad (38)$$

again in the limit that $a_{max} \gg 1$. This gives $\gamma_p b_0 T_2^* = \sqrt{(e-1)} \approx 1.3$ and $b_{RMS} \approx \sqrt{2} b_0$ for the exponential function in this limit. The exponential model is much better suited to modeling $FP(t)$ signals, as we will see when we examine some FP data below.

## A constant-gradient rectangular-vessel model

If inhomogeneities in the B-field arise mainly from the laboratory environment, then one might expect that $B(x)$ would be well described by a constant $B_0$ (provided by the main QC coil and the surrounding environment) plus an added constant field gradient across the sample (provided mainly by the environment). A constant gradient is the next-lowest-order term in an expansion of $B(x)$, and this model will be especially useful when we apply a known gradient along $\hat{z}$, as we investigate below.

To simplify this into a one-dimensional model, assume a rectangular sample vessel with $B_z(z) = B_0 + b'z$, where $b' = dB_z/dz$ is a constant field gradient aligned with one axis of the rectangular vessel. To produce a symmetrical model, we set $z = 0$ at the center of the sample vessel.

Because both $dN/dz$ and $dB_z/dz$ are constant in this model, the distribution function must be constant as well. Taking $a = b/b_{max}$, where $b_{max}$ is the maximum value at the edge of the rectangular vessel, we define $D(a) = 1$ and perform the integral in Equation (37) to obtain

$$FP(t) \sim \text{sinc}(\xi) = \frac{\sin(\xi)}{\xi} \tag{39}$$

with $\xi = \gamma_p b_{max} t$. Figure 29 shows $|\text{sinc}(\xi)|$, which is appropriate for comparisons with observations of the FP envelope signal.

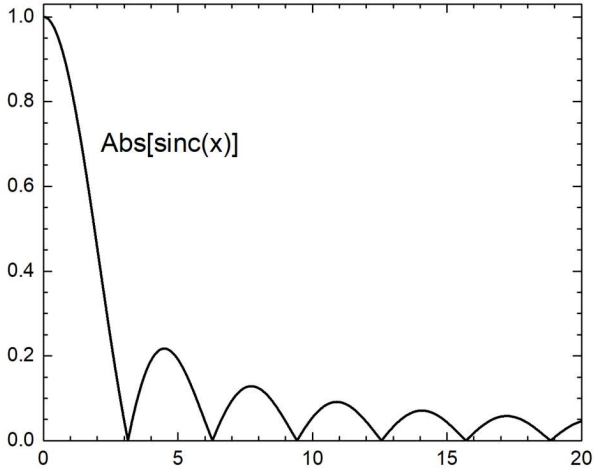

Figure 29. A plot of $|sinc(\xi)| = |sin(\xi)/\xi|$.

As seen in Figure 29, the $\text{sinc}(\xi)$ function has its first zero at $\xi = \pi$, giving $FP(T_0) = 0$ when $\gamma_p b_{max} T_0 = \pi$. Put another way, observing a first zero in $FP(t)$ (if one exists) at $t = T_{0,sec}$ (in seconds) tells us $b_{max} = (11.75/T_{0,sec})$ nT and the B-field gradient is equal to $b' = b_{max}/R$, where $R = L/2$ is the effective radius of the sample vessel (about 2 cm for the QC vessel).

The spin-vector illustration Figure 28 gives a nice illustration of why these zeros happen in the constant-gradient model. At $t = 0$, the spins are all aligned and $FP(t) = 1$. When $\gamma_p b_{max} t = \pi$, the collection of spin vectors fills the circle uniformly and we have $FP(t) = 0$. This happens again at $2\pi, 3\pi$, etc., giving the series of zeros seen in Figure 29.

## Comparisons with observations

To produce a clean measurement of the dephasing signal $FP(t)$, we found it advantageous to use the spin-echo (SE) signal as a surrogate for the initial FP decay, as this avoids transient signals associated with the detection system that mask the early-time behavior of the FP decay. Instead, the SE peak shows a clean rise and fall that appears to reproduce the initial FP signal with good fidelity. As described above, we typically correct the data for irreversible $T_2$ dephasing, because $T_2^* \ll T_2$ and we have ignored $T_2$ dephasing in our models. This correction is fairly minor, and we also normalize the data so the peak height is equal to one, again to simplify the models.

Figure 30 shows a best-case spin-echo peak taken with our QC apparatus, showing the longest effective $T_2^*$ we have observed to date, displayed in a semi-log plot to better show the tails of the dephasing signal. As shown in the figure, sum of two Lorentzians provides quite a good fit to the observations using

$$SE(t) = \frac{a_1}{1+(t/t_1)^2} + \frac{a_2}{1+(t/t_2)^2} \tag{40}$$

with $a_1 = 0.8$, $t_1 = 0.13$ seconds, $a_2 = 0.2$, and $t_2 = 1.2$ seconds for the data in Figure 30. Plugging this into the exponential model above gives the distribution of magnetic fields

$$D(b) \sim a_1 \exp(-|b\gamma_p t_1|) + a_2 \exp(-|b\gamma_p t_2|) \tag{41}$$

which is shown in Figure 31, indicating an RMS field of about 39 nT. Overall, we have found that the exponential model fits our data reasonably well, provided we find a spot where the ambient laboratory fields are not especially strong.

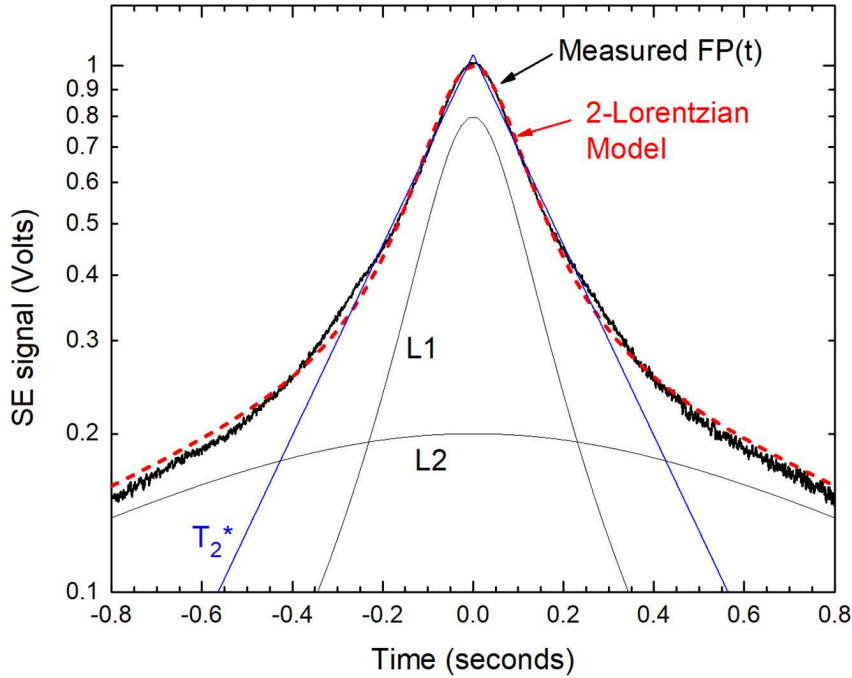

*Figure 30. This somewhat busy plot shows a spin-echo peak (black noisy line) having the longest $T_2^*$ we have observed. The blue line labeled $T_2^*$ shows a simple exponential decay $exp(-t/T_2^*)$ with $T_2^* = 0.24$ seconds. The red dotted line shows a 2-Lorenztian model of the data (described in the text), and the lines denoted L1 and L2 show the two individual Lorentzians in this model. The SE data were corrected for background noise and $T_2$ dephasing (both fairly minor corrections). These data may represent an "instrument limit" for our QC apparatus, or the field inhomogeneities may result entirely from the lab environment.*

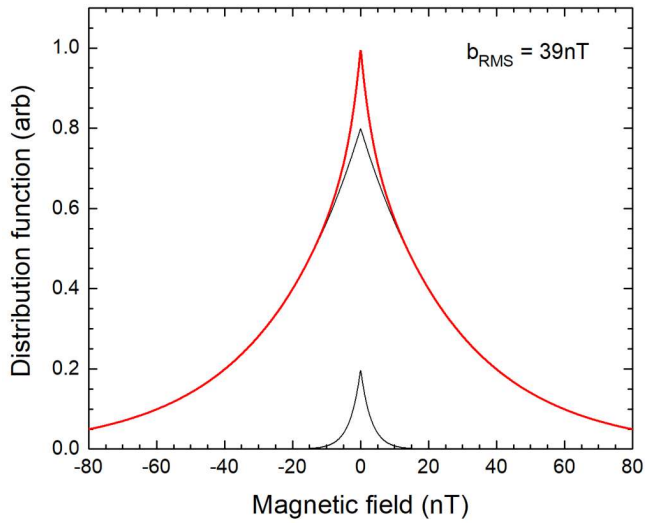

*Figure 31. The magnetic distribution function derived from our 2-Lorentzian model of the dephasing data in Figure 30. The red curve shows the total D(b), while the other curves show contributions from the two exponential functions. Although L2 is small, it contributes significantly to the slowly decaying tails in FP(t).*

## Adding external magnetic fields

The FP dephasing models becomes somewhat more interesting when we add external nonuniform magnetic fields to the picture. The QC apparatus does not include this capability, but we have found that a simple coil of wire placed on the bench underneath the QC main coil is all that is needed to produce some thought-provoking results. We typically use the 15-cm coil described above, but nearly any handy coil should produce similar behaviors.

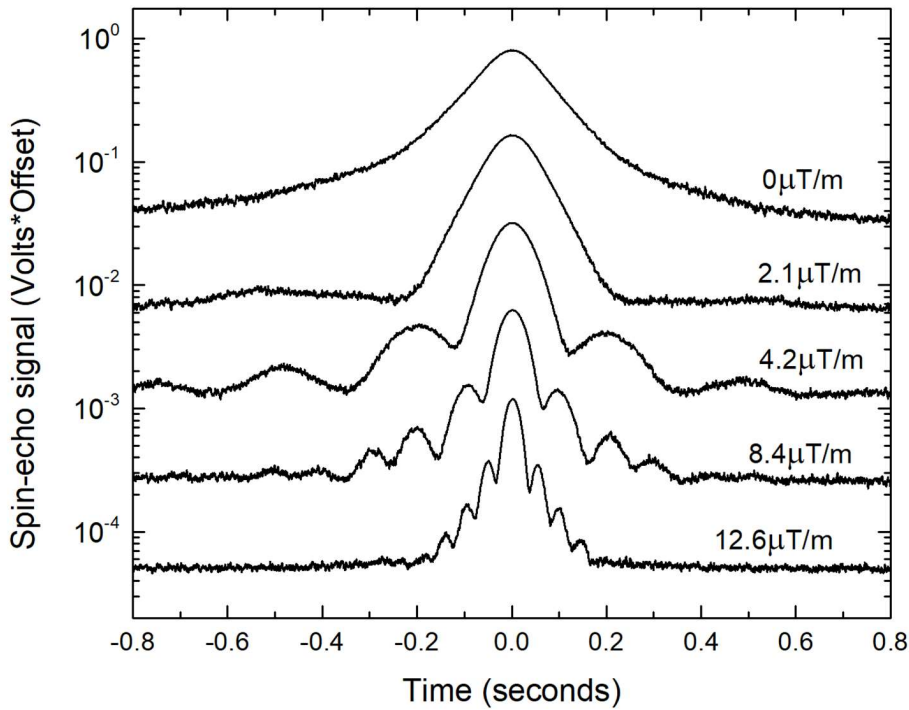

*Figure 32. This series of observations shows spin-echo signals with externally applied magnetic field gradients. The different curves are offset for clarity, and the zero-gradient curve is essentially the same as in Figure 30. The "wiggles" are roughly described by the model calculations in Figure 33, which included no adjustable parameters other than the overall normalization. In all cases, the $B_0$ current was adjusted to produce a non-oscillating mixer signal, so the lowest-order effect of the coil was to add a field gradient along the z axis. Higher derivatives of $B(z)$ are not included in the model, so the data and models do not agree in every detail.*

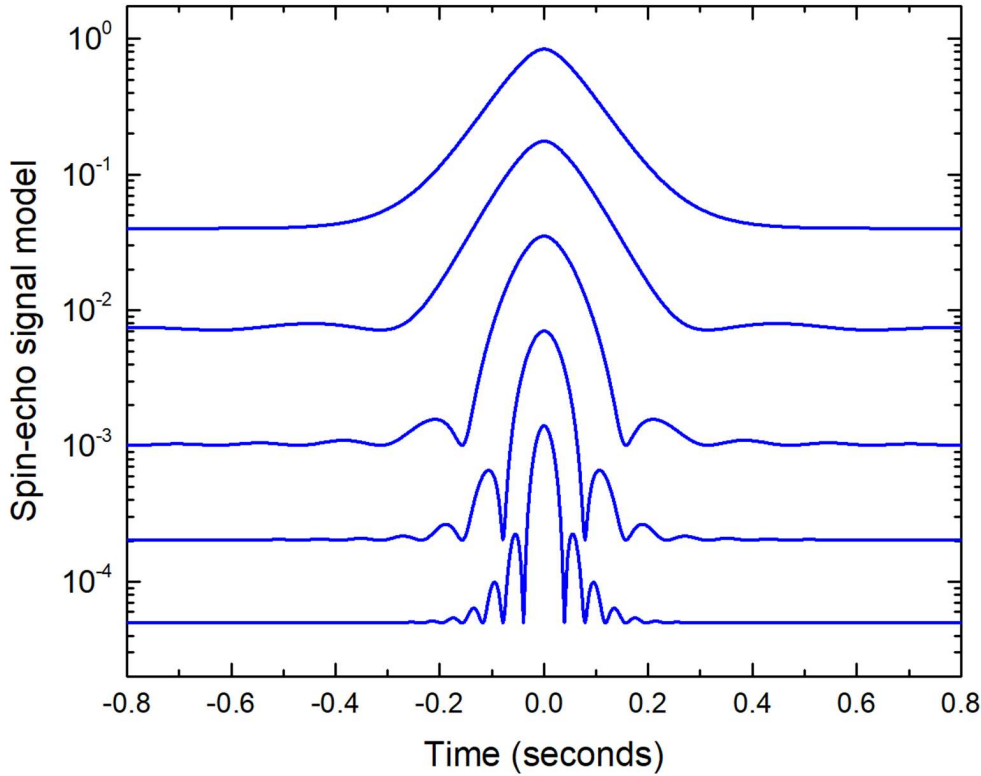

*Figure 33. This series of models corresponds to the observations shown in Figure 32, using the known applied field gradients.*

Figure 32 shows a series of observations of spin-echo peaks taken using different applied external magnetic fields gradients, and Figure 33 shows a series of constant-gradient models of the data. Other than an overall normalization factor, the models included no adjustable parameters, as the applied field gradient was known from Equation (29). The models reproduce the data in a general way, although many details are incorrect. At low applied gradients, the model does not take into account the ambient laboratory fields, and at high gradients the model does not incorporate higher-order terms in the applied coil field. Perhaps one could supply more carefully controlled external fields and get better model agreements. But doing this would distract from the simplicity of the single-external-coil experiment, which has some pedagogical value.

Figure 34 shows a more detailed view of an SE peak taken with an applied field gradient of 12.6 $\mu$T/m, together with a constant-gradient model. The model does a remarkable job reproducing the central lobe in the data while failing badly on the smaller lobes. This has been a recurring theme in our model-building efforts, and it appears that the higher-order lobes are sensitive to higher-order terms in the magnetic-field structure. In particular, we suspect that an asymmetrical model may be needed to better fit these higher-order features, which would substantially complicate the analysis. We did not pursue this investigation much further, as that would likely go too far beyond what is appropriate for the undergraduate teaching lab.

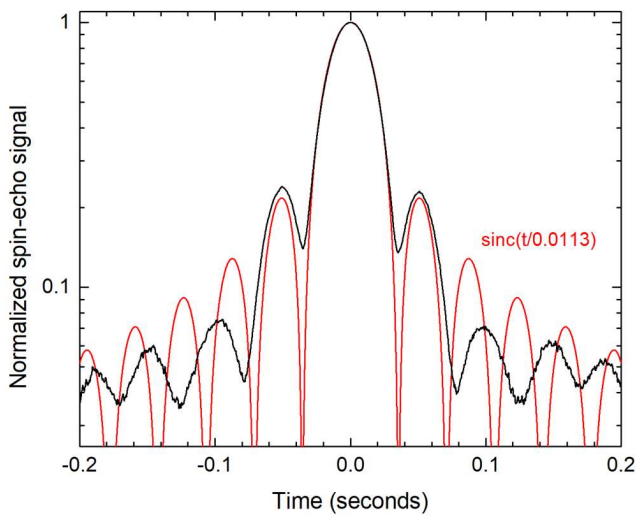

*Figure 34. Another SE run with an applied field gradient of 12.6 µT/m (black), along with a sync function fit to the SE peak (red). The model does an excellent job reproducing the central lobe but badly misses on the higher-order lobes. The positions and amplitudes of the smaller lobes changes substantially with changes in ambient laboratory fields, probably reflecting higher-order terms in the magnetic-field structure in the water sample.*

Soon after we purchased our QC apparatus, we observed SE peaks that sometimes exhibited rather complex structures on the oscilloscope, similar to those shown in Figure 35. The origin of these structures was immediately quite mysterious, as they changed with the location and orientation of the QC instrument. In one lab, the bumps and wiggles were quite pronounced, apparently because of a large rooftop air-handling system directly above the lab.

After we added our external coil and began measuring and modeling its effects, we soon realized that small ambient vertical magnetic gradients explained most of what we were seeing. We now find that Figure 35 serves as a handy guide for understanding the ambient fields that surround the QC apparatus, so this modeling exercise had an unintended benefit in this regard.

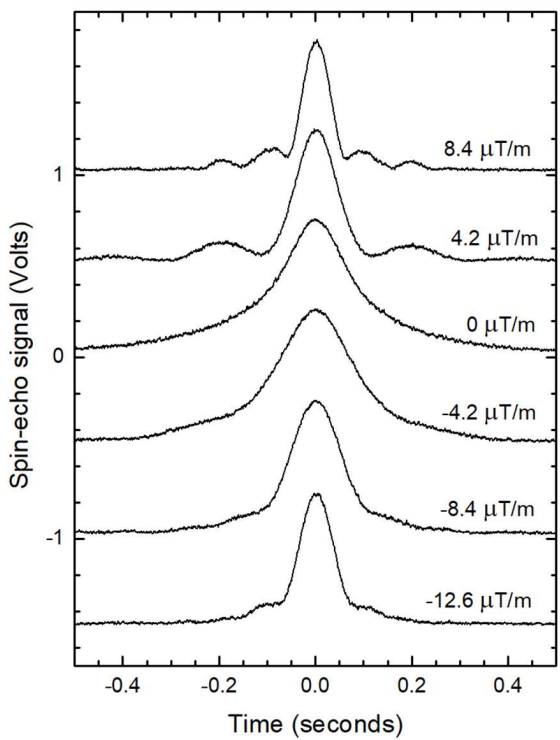

*Figure 35. This series of observed SE peaks on a linear scale illustrates what might be seen on the oscilloscope in different laboratory environments. If the ambient lab fields are weak, then the SE peak will probably be roughly Lorentzian in shape with little complex structure. If the ambient lab fields are strong, various bumps and wiggles may appear in the data.*

## Spin Echo dephasing

The spin-echo phenomenon can only happen if the initial dephasing of the FP signal is reversible, caused by (typically very small) nonuniformities in the $B_0$ magnetic field. As long as these nonuniformities are fixed in time, flipping the spins with a $\pi$ pulse will reverse the dephasing and produce a spin-echo peak. Not all dephasing is reversible, however, and we now examine how different physical processes can reduce the amplitude of the spin-echo peak, yielding easily measurable signals in the QC apparatus.

The most fundamental source of irreversible dephasing comes from nanoscopic interactions between water molecules. The coherent precession of the proton spins is a non-equilibrium phenomenon, so molecular interactions will transfer energy from this motion to the thermal bath, eventually producing a completely random spin distribution that produces no coherent FP signal. The result is a friction-like phenomenon – difficult to derive from first principles, but easy to incorporate into the equations of motion using a phenomenological constant. This is done in the Bloch equations, giving the dephasing time constant $T_2$. Like the dipole relaxation time $T_1$, the value of $T_2$ depends only on the intrinsic molecular properties of the sample and is independent of the applied magnetic fields.

## Measuring $T_2$

Assuming the spin-echo process completely reverses $T_2^*$ dephasing (a reasonable assumption in some circumstances), we can determine $T_2$ simply by measuring the height of the spin-echo peak in comparison to the initial FP signal as a function of the pulse separation time $\tau$. Figure 36 shows some especially clean data we obtained after letting the instrument warm up for 24 hours. The top curve in this figure shows a simple exponential decay $SE_{peak}(t) = S_0 \exp(-t/T_2)$ with $S_0 = 1765$ mV and $T_2 = 2.73$ seconds. The lower curve adds an additional decay factor to better fit the data with $SE_{peak}(t) = S_0 \exp(-t/T_2) \cdot \exp[-(t/T_3)^3]$, with $T_3 = 5.82$ seconds.

The $T_2$ decay factor in water is well known, although we cannot calculate either $T_1$ or $T_2$ from our limited understanding of complex molecular interactions. Both numbers also depend on temperature and chemical impurities, especially paramagnetic salts and other solutes that can dramatically shorten both decay times. We used grocery-store deionized water in our measurements, with no attempt to keep our sample especially pure. One finds $T_2 < T_1$ in many liquids, and this is our finding in water as well.

## A diffusion model

The additional $\exp[-(t/T_3)^3]$ factor can arise from several physical mechanisms, and this particular functional form may not be accurate over all time periods. The spin-echo process assumes that all magnetic fields are constant in time, and that all the spins are precisely fixed in space. Any mechanisms that violate these assumptions could cause the SE signal to decay faster than expected from $T_2$ processes alone.

One well-known mechanism that might explain the behavior seen in Figure 36 is molecular diffusion in the sample. The basic idea in this model is that each spin randomly diffuses during the spin-echo process, which changes its precession frequency because of spatial inhomogeneities in the magnetic field. Because of these changes, the spin-echo reconstruction of the initial magnetic moment will be disrupted, yielding a smaller spin-echo peak. The longer the total precession time, the greater the disruption, and thus the smaller the spin-echo peak.

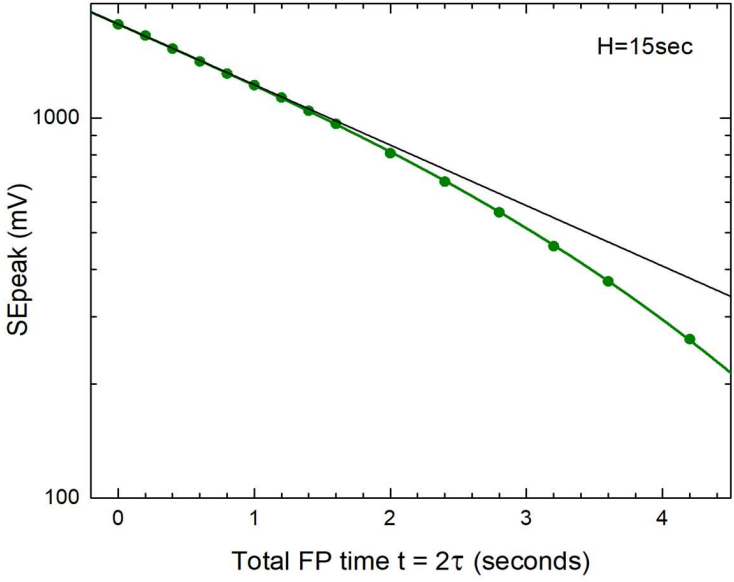

Figure 36. Data points in this plot show measurements of the height of the spin-echo peak as a function of the free-precession time (equal to twice the pulse delay time τ). The top curve shows a simple exponential decay $exp(-t/T_2)$ with $T_2 = 2.73$ seconds, and the lower curve adds an additional cubic decay factor $exp[-(t/T_3)^3]$ with $T_3 = 5.82$ seconds, as described in the text. The data were taken after the QC instrument had warmed up for 24 hours. With insufficient warmup, the SE peak drops off substantially faster with time.

As a rough estimate of the magnitude of this effect, consider that normal diffusion causes each spin to execute a random walk through the sample, moving it an average distance of $\delta x \approx \sqrt{Dt}$ over a time $t$, where $D \approx 2.3 \times 10^{-9} m^2 s^{-1}$ is the diffusion constant in water at 25C ($D \approx 3.2 \times 10^{-9} m^2 s^{-1}$ at 40C) [2000Hol]. If we assume a constant magnetic field gradient throughout the sample, then the change in magnetic field during this time is $\delta b \approx (\partial B_0/\partial x)\sqrt{Dt}$, which causes the overall precession angle $\varphi$ to wander by $\varphi = (\gamma_p \delta b) t$, thus giving

$$\varphi^2 \approx \gamma_p^2 \left(\frac{\partial B_0}{\partial x}\right)^2 Dt^3 \tag{42}$$

As with the free-precession signal, this dephasing causes the signal to diminish by $\sim \cos(\varphi)$ (in the ensemble average). In the limit of small $\varphi$, we write $\cos(\varphi) \approx 1 - \varphi^2/2 \approx \exp(-\varphi^2/2)$, giving a decay rate from diffusion of

$$SE_{peak}(t) \sim \exp\left[-\gamma_p^2 \left(\frac{\partial B_0}{\partial x}\right)^2 \frac{Dt^3}{2}\right] \tag{43}$$

Replacing this rough estimate with a more rigorous ensemble average yields the result [1954Car]

$$SE_{peak}(t) \approx S_0 \exp\left[-\frac{t}{T_2}\right] \exp\left[-\gamma_p^2 \left(\frac{\partial B_0}{\partial x}\right)^2 \frac{Dt^3}{12}\right] \tag{44}$$

In the absence of a strong applied B-field gradient, the magnetic field inhomogeneities in the QC water sample are probably not well described by a constant gradient model (as we found in our FP dephasing analysis), so this diffusion model is likely only a rough approximation of the full system.

Ignoring these issues, comparing Equation (44) with the data in Figure 36 gives

$$\frac{\partial B_0}{\partial x} \approx \sqrt{\frac{12}{T_3^3 \gamma_p^2 D}} \approx 17 \; \mu T/m \qquad (45)$$

This result is not totally unreasonable, given that the constant-gradient theory that yields Equation (44) is a bit simplistic for our situation. However, our FP analysis above suggests $(\partial B_0/\partial z) < 4 \mu T/m$ with no applied gradients and low ambient fields, so 17 $\mu T/m$ is certainly somewhat higher than expected.

## A convection model

While investigating this phenomenon, we soon found that $SE_{peak}(t_{max})$ at $t_{max} = 2\tau_{max} = 4.2$ seconds was quite sensitive to various experimental parameters, and this provides some additional clues about the dominant physical mechanisms responsible for the observed SE dephasing. For example, Figure 37 shows $SE_{peak}(t_{max})$ as the QC instrument warmed up, both with the top cover on and off. This peculiar behavior cannot be explained by the basic diffusion model, and we soon realized that convection in the water vessel was probably providing an additional transport mechanism. Moreover, in many circumstances it appears that convection is more important than diffusion in the SE decay process.

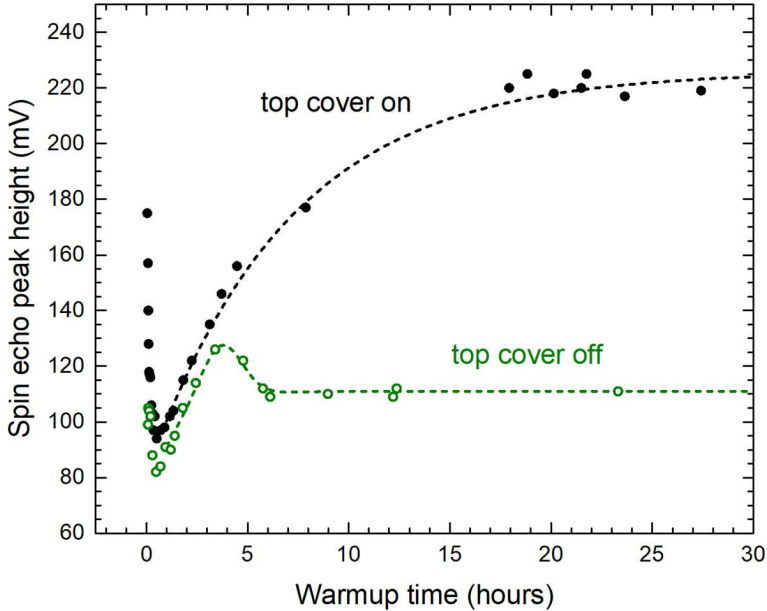

Figure 37. These data show the height of the spin-echo peak with τ=2.1 seconds as the QC warms up after being turned on at t=0. The black data points were taken with the top cover on, which is the normal QC configuration. Removing this cover in a separate run gave the green points. In both cases, lines were drawn through the data to guide the eye. As described in the text, a convection model seems to best describe these behaviors.

Although convective transport is difficult to model, a convection model can qualitatively explain all the behaviors seen in Figure 37. Focusing first on the "cover-on", data, we see that $SE_{peak}(t_{max})$ is relatively large when the instrument is first turned on, when the vessel and its surroundings are all essentially in thermal equilibrium. With no temperature gradients in the system, there will be little water convection, so the spin-echo phenomenon yields a large peak. Shortly after turn-on, however, the main coil begins to warm up, producing thermal gradients that soon drive convection in the water vessel. And this causes the SE peak to drop quickly, reaching a minimum value after about 30 minutes. At later times, warm air is trapped in the interior space of the main coil, so the interior temperature gradients lessen, yielding a slowly increasing SE peak. The top-cover approached its maximum temperature of ~40 C after some hours, causing the SE peak to finally reach a stable maximum after about 15 hours.

With the top cover removed, the green data points in Figure 37 show roughly the same initial SE peak drop-off as with the cover on, as one would expect. With no top cover to trap the warm air in the main coil tube, however, air convection prevents the formation of a stably stratified column of air, thus (in our interpretation of the data) yielding strong temperature gradients around the water vessel that do not diminish over time. As a result, convection in the water vessel remains strong at later times, reducing the $SE_{peak}(t_{max})$ accordingly. Throughout both runs, we also occasionally monitored the $SE_{peak}$ at $t = 1$ second, finding that it was remarkably stable, never varying by more than a percent or two.

Because that the data in Figure 36 were taken after the QC instrument warmed up for over 24 hours, this run gave the highest observed $SE_{peak}(t_{max})$ and the most stable data overall. In contrast, taking data after a short warm-up period yielded somewhat erratic results and smaller values of $T_3$, suggesting that convection is the dominant mechanism when the QC apparatus is not fully warmed up.

Although we have not found a quantitative model of how convection affects the SE peak, we can at least estimate the velocity of convective motion. In the inertial drag regime, the terminal velocity of a rising parcel of warm water will be of order

$$v_{conv} \approx \sqrt{\frac{8gR\lambda \cdot \delta T}{3C_d}} \qquad (46)$$

where $\lambda \approx 2 \times 10^{-4}/°C$ is the thermal-expansion coefficient of water, $g$ is the gravitational acceleration, $R$ is the convective eddy size, $\delta T$ is the excess temperature of a rising parcel, and $C_d \approx 0.5$ is the drag coefficient.

In the diffusion model described above, a water molecule random walks roughly according to $\delta x \approx \sqrt{Dt}$, giving $\delta x \approx 100$ $\mu$m in 4 seconds. Assuming a parcel size of $R \approx 1$ cm, Equation (46) gives a similar motion when $\delta T$~10 $\mu$K. Thus even this order-of-magnitude estimate suggests that convective transport can be quite large compared to diffusive transport, confirming our conclusion that convection may be the dominant cause of the observed enhanced dephasing of the SE peak.

Again with the QC apparatus fully warmed up, we further observed how $SE_{peak}(t_{max})$ changed with the application of an external magnetic field gradient. We used the 15-cm coil described above to produce a known $B' = dB/dz$ in the sample vessel while always adjusting $B_0$ to maintain a near-constant mixer signal.

The diffusion model in Equation (44) predicts $SE_{peak}(t_{max}) \sim \exp[-(B'/B_2)^2]$ with $B_2 \approx 27$ µT/m, and the data in Figure 38 show reasonable agreement with this model. We would expect that a convection model would likely yield a similar functional form, as both diffusion and convection involve particle transport in a non-uniform magnetic field during the spin-echo process.

Our overall conclusion from these experiments is that it appears that both diffusion and convection are likely important in explaining the observed $SE_{peak}(t)$ behaviors at large $t$. Convection almost certainly dominates when the QC is warming up, but diffusion may become the dominant mechanism once the system is fully stabilized. It is difficult to separate the two mechanisms using the QC data alone.

In the undergraduate teaching lab, students should be able to measure $T_2$ with reasonable accuracy even with modest warmup times. Because the $\exp[-(t/T_3)^3]$ factor depends so strongly on time, the SE peak data with $t < 2$ seconds typically provides a clear $\exp(-t/T_2)$ signal. Measuring both $T_1$ and $T_2$ nicely segues to a discussion of MRI technology, as medical MRI machines typically focus on proton precession signals using measurements of $T_1$ and $T_2$ to produce image contrast, as both these quantities are sensitive to the chemical makeup of the sample. The transport issues encountered while measuring $T_2$ serve as a pedagogically valuable reminder that unwanted effects can easily produce a host of systematic errors in physical measurements.

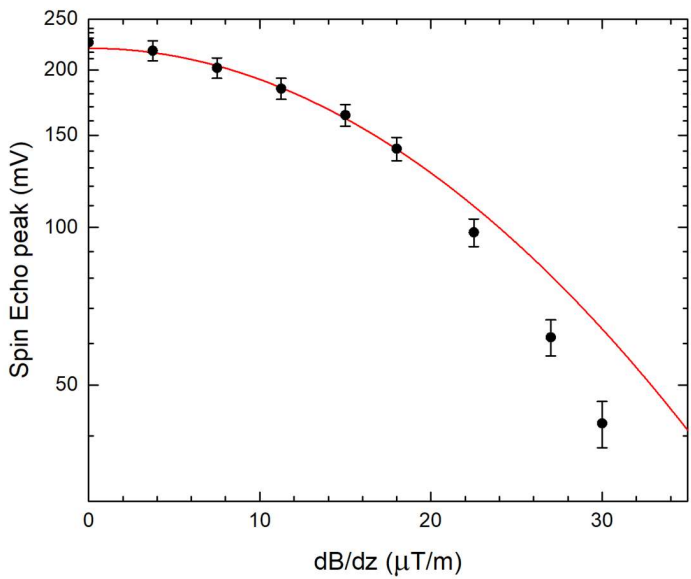

Figure 38. These data show measurements of the spin-echo peak height as a function of an applied external B-field gradient, using $H = 15$ seconds and $\tau = 2.1$ seconds. The line shows $SEpeak(t) = SE_0 exp[-(B'/B_2)^2]$ with $B_2 = 27$ µT/m.

## Summary

As we mentioned at the beginning of this article, NMR has long been a staple in physics undergraduate teaching labs, as the topic incorporates so many fundamental physics concepts and techniques in a relatively simple apparatus. TeachSpin's *Quantum Control* instrument hits a definite sweet spot in this area, with its easy-to-use microprocessor interface, beautifully engineered hardware, and modest price tag. We have found that this instrument is a delight to work with, providing remarkably clean and stable measurements with high signal-to-noise ratios. Plus the underlying physics is fascinating while remaining quite accessible over a range of educational levels.

# Acknowledgements

This work was supported in part by a generous donation from Beatrice and Sai-Wai Fu to the Physics Teaching Labs at Caltech, together with Caltech's long-standing support of outstanding laboratory instruction across many STEM fields.

*Contact:* For corrections, comments, or just to compare notes, please contact Kenneth G. Libbrecht, *klibbrecht@gmail.com*, or mail to: Mail-stop 264-33 Caltech, Pasadena, CA 91125.

# Appendix 1 – Modeling the FP noise background

To better understand the noise in the QC free-precession signal, we first note that the detector coil is part of a tuned circuit that acts much like a narrow-band electronic filter (as described in the TeachSpin QC manual), reducing the bandwidth of the signal and its accompanying electronic noise. Figure 39 shows a measurement of the coil response, which we measured by placing an external coil near the QC apparatus and doing a frequency-response analysis (FRA) measurement of the FP signal as a function of the external-coil drive frequency.

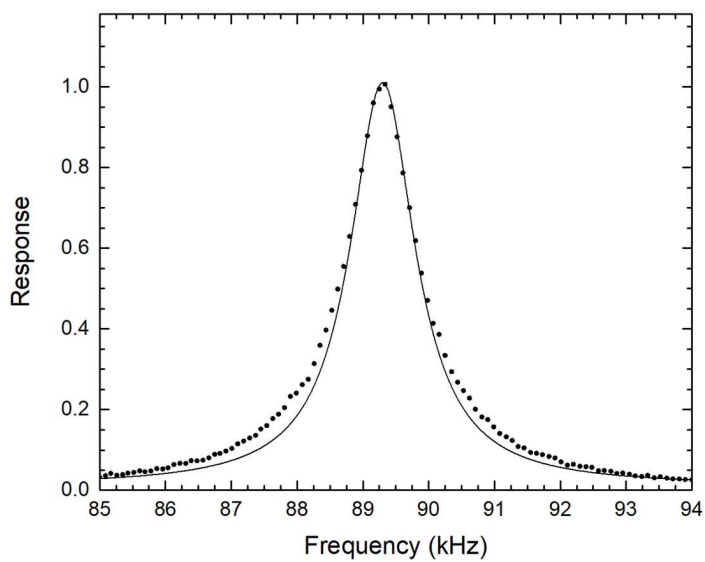

*Figure 39. The data points in this plot shows the measured narrow-band response of the detector coil when driven by the magnetic field from an external coil placed outside the QC apparatus. No NMR signal was present during this response measurement. The line shows a Lorentzian curve centered at 89.3 kHz with a half-width of 0.6 kHz.*

Given this high-Q bandpass filter, we model the noise as a sinusoidal signal around 89.3 kHz with a slowly changing amplitude and phase. Assuming Gaussian statistics, we model the spectral density of the amplitude noise with

$$A(x, y) \sim \exp[-(x^2 + y^2)/\sigma^2] \quad (47)$$

with the two axes represent the $A_x$ and $A_y$ components of the sinusoidal noise voltage. With this noise model, averaging many oscilloscope traces of the noise-only QC envelope signal gives the average background noise

$$FP_{noise} = \frac{\int_{-\infty}^{\infty} \int_{-\infty}^{\infty} \sqrt{x^2 + y^2} \exp[-(x^2 + y^2)/\sigma^2] dx dy}{\int_{-\infty}^{\infty} \int_{-\infty}^{\infty} \exp[-(x^2 + y^2)/\sigma^2] dx dy} = \frac{\pi^{3/2} \sigma^3 / 2}{\pi \sigma^2} = \frac{\pi^{1/2}}{2} \sigma \quad (48)$$

If we include an NMR signal of amplitude $a$ along the $\hat{x}$ direction, then the numerator in the above expression is replaced with

$$\int_{-\infty}^{\infty} \int_{-\infty}^{\infty} \sqrt{(x+a)^2 + y^2} \exp[-(x^2 + y^2)/\sigma^2] dx dy \quad (49)$$

and this integral cannot be expressed in a simple closed form. However, evaluating Equation (49) numerically gives the results shown in Figure 40, which shows our model of the measured signal+noise along with two methods one might use to correct the data, all normalized by $\sigma$.

From this noise analysis, we conclude that a quadratic noise correction, given by

$$FP_{quad} = \sqrt{FP_{meas}^2 - FP_{noise}^2} \quad (50)$$

gives a reasonable estimate of the actual signal amplitude, where here $FP_{meas}$ is the direct envelope signal and $FP_{noise}$ is the average background signal. At the same time, using a naïve correction of $FP_{corr} = FP_{meas} - FP_{noise}$ gives numbers that are even worse than using the uncorrected data. For most of the measurements in this document, therefore, we replaced $FP_{meas}$ with $FP_{quad}$ to reduce systematic errors that otherwise arise from the noise.

Of course, this noise correction is only as good as our noise model, and Figure 41 illustrates that the model does have some deficiencies. Given the noise statistics and the noise analysis above, we conclude that the $FP_{quad}$ corrected signal should provide a reasonable representation of the actual free-precession signal at perhaps the few-millivolt level, which is certainly better than using the raw measurements. Larger systematic errors could be present as well from unknown sources, as is true of any experimental apparatus.

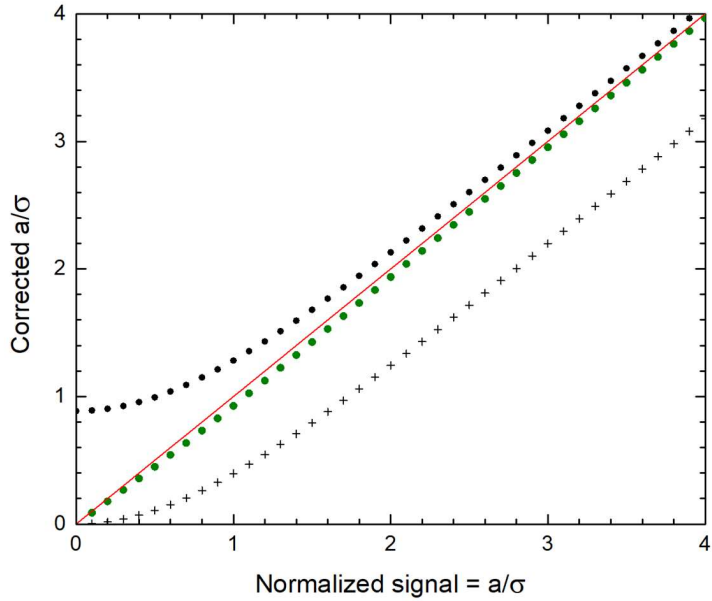

*Figure 40. The red line in this plot shows the FP envelope signal level one wishes to measure, normalized by the noise parameter σ. The upper points (black dots) show the measured $FP_{meas}$ signal, which includes noise. The + symbols show ($FP_{corr} = FP_{meas} - FP_{noise}$), revealing that this naïve noise subtraction introduces large systematic errors in the corrected measurements. The middle points (green dots) show the quadratic noise correction described in the text, showing that this correction method provides a reasonable estimate of the actual signal level. In this noise model, the corrected points are lower than the actual signal by at most about five percent, which is certainly better than using no noise correction.*

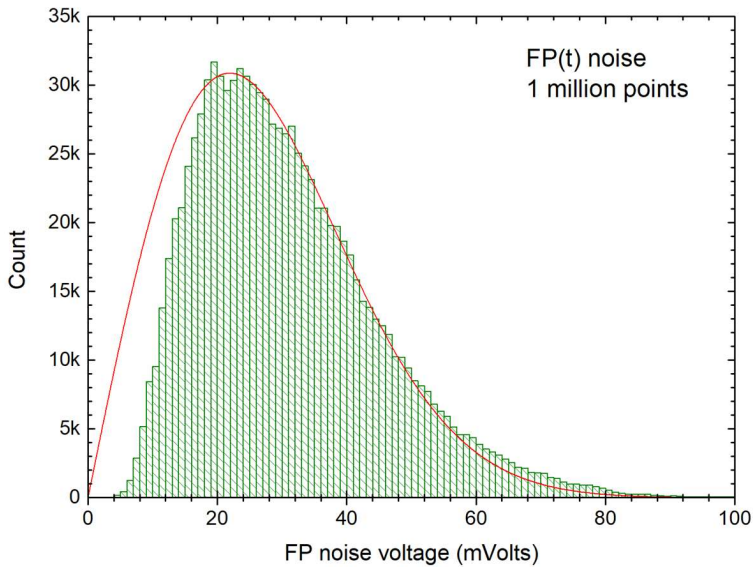

*Figure 41. This graph shows noise statistics we measured using our QC apparatus. With zero NMR signal present, we recorded the FP envelope signal for several seconds and created a histogram of the measured signal values. The data show that the noise is positive definite with an average noise signal of about 30 mV, which is easily verifiable from a simple oscilloscope trace. The red line shows a prediction from our noise model, indicating that the real data are lacking low-voltage values. This discrepancy could result from unmodeled instrumental offsets or any number of unknown factors.*